\documentclass[aps,prd,preprintnumbers,superscriptaddress,nofootinbib]{revtex4}

\usepackage{hyperref}
\hypersetup{colorlinks=true,linkcolor=purple,anchorcolor=blue,citecolor=blue, filecolor=blue,urlcolor=red,bookmarksnumbered=true,
pdfview=FitB
}
\usepackage{color}
\usepackage{xcolor}
\colorlet{purple1}{blue!70!red}
\colorlet{darkred}{red!50!black}

\usepackage{graphicx}
\usepackage[bf,SL,BF]{subfigure}
\usepackage{psfrag}
\usepackage{color}
\usepackage{amssymb}
\usepackage{amsmath}
\usepackage{epstopdf}
\usepackage{bbold}
\usepackage[normalem]{ulem}

\newcommand{\be}{\begin{eqnarray}}
\newcommand{\ee}{\end{eqnarray}}

\begin{document}
\title{
Empirical formula for total inelastic cross-section of proton-nucleus scattering
}

\author{Hemant Kumar}
 \affiliation{Department of Physics, National Institute of Technology Kurukshetra, India 136119}
 
\author{Tanmay Maji }
\thanks{Corresponding Author} 
\email{tanmayphy@nitkkr.ac.in}
\affiliation{Department of Physics, National Institute of Technology Kurukshetra, India 136119}

\author{Deepa Gupta}%
\affiliation{Department of Physics, National Institute of Technology Kurukshetra, India 136119}
\affiliation{Indian Institute of Technology Roorkee, Uttarakhand, India 247667 }

\author{Ashavani Kumar}
 \affiliation{Department of Physics, National Institute of Technology Kurukshetra, India 136119}

\date{\today}

\begin{abstract}

We propose a generic empirical formula for total inelastic cross-sections for various target nuclei scattered by a proton at different energies, which is applicable over a wide range of energy from $15 ~MeV$ to $1~ TeV$. The proposed model is parameterized based on the fitting of extensively studied experimental cross-section data for the Aluminium and Carbon nucleus targets, considering factorization over high-energy and low-energy regimes. The parameters in high-energy formula are determined by the fitting of the high-energy saturation value of the inelastic scattering cross-section data with mass numbers. The universality of the empirical formula is investigated by comparing the model prediction with the experimental data of inelastic proton-nucleus scattering over a wide range from light elements such as Deuterium to heavy elements such as Uranium. A detailed comparison with the existing models and GEANT4 simulation is also presented.

\end{abstract}

\maketitle


\section{\label{sec:level1}Introduction}

Proton–nucleus interactions provide a fundamental testing ground for our understanding of hadronic reactions in the nuclear medium. Among the key observables characterizing such interactions is the proton–nucleus inelastic cross section, defined as the total cross section minus the elastic contribution. This quantity encodes the probability for all inelastic processes, including particle production, nucleon emission, and nuclear excitation etc. Inelastic cross-sectional data play a fundamental role in medical radiation therapy \cite{newhauser2015physics}, accelerator shielding and design\cite{chen2005calculations}, space radiation protection, detector development, and cosmic rays\cite{silberberg1973partial}. 

Several recent experimental efforts have been devoted to improve our understanding of proton inelastic scattering cross sections across a range of target nuclei and energy regimes. Measurements of proton-induced inelastic transitions on $^{56}$Fe have been conducted using high-resolution $\gamma$-ray detection at a tandem accelerator, which provides detailed cross-section data for the strongest inelastic channels \cite{COMAN2025123080}. 
While for light nuclei such as $^{16}$O and $^{28}$Si, precise $p,p'$ cross sections are also measured experimentally \cite{PhysRevC.101.024604}.
Additionally, proton inelastic scattering has been used as a sensitive probe of nuclear structure in exotic systems; for instance, inelastic scattering off neutron-rich helium isotopes has revealed deformation effects and provided benchmarks for microscopic nuclear theories \cite{PhysRevC.106.024609}. These experiments not only provide support for the development and validation of global reaction models such as TALYS \cite{Koning2023TALYS}, but also provide important input for the design of new nuclear facilities. 
A reliable knowledge of proton–nucleus inelastic cross sections over a broad range of energies and target masses has drawn attention for both phenomenological studies and the validation of reaction models.

At high incident energies, proton–nucleus reaction cross sections exhibit a relatively smooth dependence on the target mass number and approach asymptotic values governed primarily by nuclear geometry. Early experimental systematic demonstrated that, above a few GeV, the cross sections become nearly energy independent and scale approximately with the nuclear radius, with deviations at the level of a few percent [1]. These observations motivated the development of simple empirical parameterizations capable of reproducing the global trends of the data.
The first models for the total inelastic cross section $\sigma(A)$ were based on simple scaling laws with the atomic number $A$ as
\begin{equation}
    \sigma_{PL}(A)=44.9A^{0.7},
    \label{eq:1}
\end{equation}
which provides reasonable estimates for some of the nucleus, but lack precision for higher values of $A$. This model is known as the power law (PL) model.

Letaw $\textit{et. al.}$ \cite{letaw1983} proposed one of the earliest and simplest empirical models to estimate the total inelastic cross section for proton–nucleus scattering, valid over the energy range of 10 MeV to 1 GeV. The model was based on a large compilation of experimental data and provides separate expressions for low-energy and high-energy regions. Their parameterization successfully reproduced the asymptotic high-energy behavior while also accounting for the characteristic energy dependence observed at intermediate and low energies. In particular, their analysis revealed a shallow minimum in the cross section near a few hundred MeV, followed by a pronounced rise toward lower energies, attributed to the increasing importance of nuclear interaction effects. Despite its simplicity and widespread use, the accuracy of this formulation deteriorates for light nuclei and in energy regions where nuclear medium effects become significant.
At higher proton energies, the cross section depends only on the atomic mass number$ A$ of the target nucleus and is given by
\begin{equation}
    \sigma(h.e) = 45 \, A^{0.7} \left[1 + 0.016 \sin(5.3 - 2.63 \ln A) \right].
    \label{eq:2}
\end{equation}
At lower proton energies, an additional energy-dependent term is introduced to account for deviations from high-energy behavior, which reads as
\begin{equation}
    \sigma(E) = \sigma(h.e) \cdot f(E),
\end{equation}
with
\begin{equation}
    f(E) = 1 - 0.62 \exp\left(-\frac{E}{200}\right) \sin(10.9 \, E^{-0.28}),
\end{equation}
where$ f(E)$ introduces the energy dependence in the low-energy region, with$ E$ being the kinetic energy of the incident proton.
While the model is compact and widely used, it fails to accurately predict cross sections below the Coulomb barrier and for heavier nuclei and low-energy region.

Subsequent effort aimed to improve empirical description is proposed by Shen\cite{shen1991}, and followed the general framework originally proposed by Letaw $\textit{et. al.}$ but introduced a correction term to account for Coulomb effects, with the aim of improving the accuracy of the model in the low-energy region. This adjustment helps the model better match experimental data where the proton's energy is near or below the Coulomb barrier, and the total inelastic cross section is presented as
\begin{equation}
    \sigma^p = 42.6 \, A^{0.701} \, f^p(A) \, g^p(E) \, h^p(A, E),
\end{equation}
where the individual components are defined as
\begin{align}
    f^p(A) &= 1 + 0.0144 \sin(3.63 - 2.82 \ln A), \\
    g^p(E) &= 1 - 0.67 \exp\left(-\frac{E}{150}\right) \sin(12E^{-0.289}), \\
    h^p(A, E) &= \frac{1}{1 + (0.018A^2 - 1.15A)E^{-2}}.
\end{align}
$f^p(A)$ adjusts the cross section based on the target nucleus mass, $g^p(E)$ introduces energy dependence, and $h^p(A, E)$ is a damping function that includes the Coulomb effect.
This model works reasonably well across many elements, but caution is needed when applying it to lighter nuclei ($A < 64$), especially at low energies where the $h^p(A, E)$ term can cause the cross section to become negative.

 The Tripathi $\textit{et. al.}$ \cite{tripathi1997} developed a universal parameterization of reaction cross-sections in which proton–nucleus collisions were treated as a limiting case of nucleus–nucleus interactions. By embedding effects into an effective interaction radius and energy-dependent correction factors, the Tripathi $\textit{et. al.}$ parameterization achieved improved agreement with experimental data across a wide range of targets and energies. However, the resulting expressions are relatively complex and involve multiple conditional forms, which can obscure the underlying physics and complicate systematic analyses.
 Upon analysis of this model, it was observed that it requires target-specific tuning and its formulation is relatively lengthy. Among all existing models, the model provides the most consistent and accurate results for light elements. However, despite its success in the low mass region, it does not produce reliable results for heavier nuclei, especially for elements $A> 64$. For a detailed explanation, see \cite{tripathi1997}.

Recently, renewed interest in precise hadronic cross sections has led to further refinement of empirical approaches. Nakano, $et. al.$ \cite{nakano2021}  proposed a new parameterization of nucleon-induced nonelastic cross sections based on physical insights obtained from improvising intranuclear cascade (INC) model analyses \cite{PhysRev.72.1114,AICHELIN198614,10.1143/ptp/87.5.1185,CUGNON1987558,PhysRevC.66.044615,PhysRevC.69.064611,CUGNON20071332,PEDOUX201116}.
Their formulation incorporates two key physical concepts: the Discrete Level Constraint (DLC) effect and the Coulomb effect. This model is designed to cover a wide energy range from 0 to 2000 MeV. It closely follows the experimental data in both the low and high energy regions without significant limitations, except for nuclei with mass numbers $A < 11$. The model provides consistent and reliable results for heavier nuclei. Although it successfully integrates important physical effects and shows excellent agreement with experimental data across a broad energy spectrum, it does not perform well for light nuclei. Additionally, the model is highly complex-it involves a large number of parameters and a lengthy mathematical formulation. Among all existing models, Nakano $\textit{et. al.}$’s approach uses the largest number of parameters \cite{nakano2021}.

Despite these advances, any existing empirical model has not achieved uniform precision across all energies and target masses while remaining computationally simple. They exhibit limitations under certain conditions, particularly for low-energy, lighter nuclei, and some of these models require target-specific tuning. This continued need for systematic studies that reassess existing parameterizations and explore possible refinements motivated by a simple computation of underlying reaction dynamics.

In this paper, we develop a new empirical formula for the proton–nucleus inelastic cross section, which provides reasonably high precision in cross-section to the atomic number dependency and is simple in computation. We planned to cover a wide energy range from 15 MeV to 1 TeV, making it suitable for both low- and high-energy applications. The model has been compared with available experimental cross-section measurement for all 33 different nuclei, including light elements such as Deuterium to heavier ones such as lead and Uranium. The model results are compared with existing well-known models \cite{letaw1983,shen1991,tripathi1997,nakano2021} as well as with GEANT4 simulation results. Since experimental data are lacking for several nuclei and energy ranges, having a reliable empirical formula can be useful to predict the unmeasured Nuclear cross-sections.

GEANT4 (GEometry ANd Tracking) is a general-purpose Monte Carlo simulation toolkit widely used for modeling particle interactions with matter across a broad energy range \cite{Geant4-2003,Geant4PhysicsReference,allison2006geant4,allison2016recent}. It provides a comprehensive framework that includes geometry modeling, tracking, detector response, visualization, and a broad set of physics models for electromagnetic, hadronic, and nuclear interaction models, enabling realistic simulations of scattering processes and reaction cross sections. The GEANT4 provides a variety of validated hadronic physics lists— such as \texttt{FTFP\_BERT}, which combine nuclear cascade models at low and intermediate energies with string-based descriptions at higher energies\cite{Grichine2009,gribov1969glauber,franco1966total}. These physics lists enable consistent simulations of proton–nucleus interactions from the MeV to multi-TeV regime and are extensively employed in collider and nuclear physics phenomenology for interpreting scattering data and bench marking reaction models \cite{Bagulya:2024nka,Ivanchenko:2017rvi}.

The paper is organized as follows: in the next section Sec.\ref{model}, a detailed description of the model construction is presented. In Sec.\ref{sec:numerical_results}, the numerical results and the predictions of the model are discussed, and conclude in Sec.\ref{conclusion}. 

\section{Model Construction}\label{model}
In this section, we present a detailed formulation of the proposed empirical model for the inelastic cross section of proton-nucleus scattering. 
This approach is inspired by the spirit of the widely used Letaw $\textit{et. al.}$ model~\cite{letaw1983}, while incorporating updated experimental information and additional physical considerations. The inelastic cross-section for the wider range of proton energy shows a classified scale dependency, hence factorized in high-energy and low-energy regions as follows. 
\subsubsection{High Energy Corrections}
\begin{figure}[h]
    \centering
    \includegraphics[width=0.37\linewidth, trim=100 240 100 240, clip]{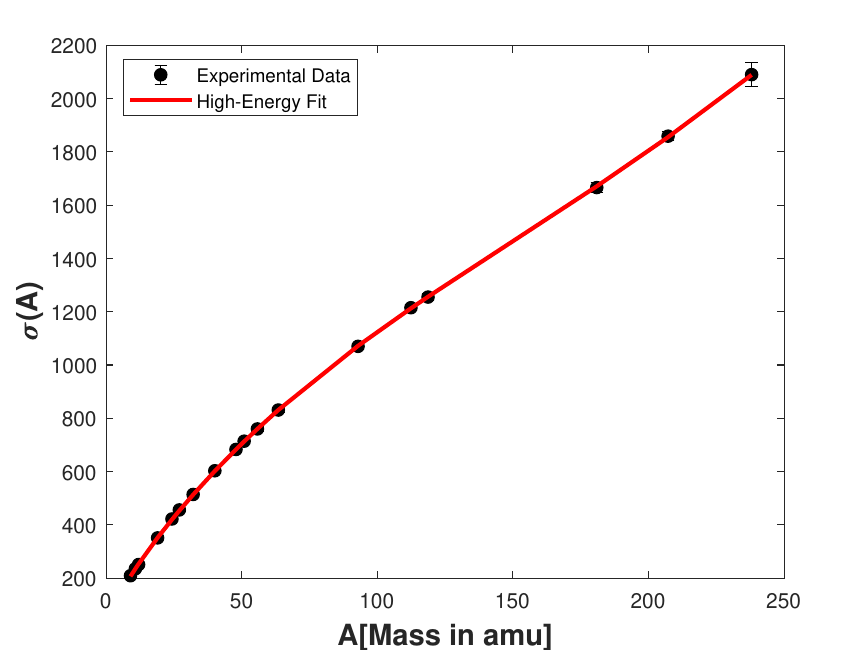}
    \includegraphics[width=.6\linewidth]{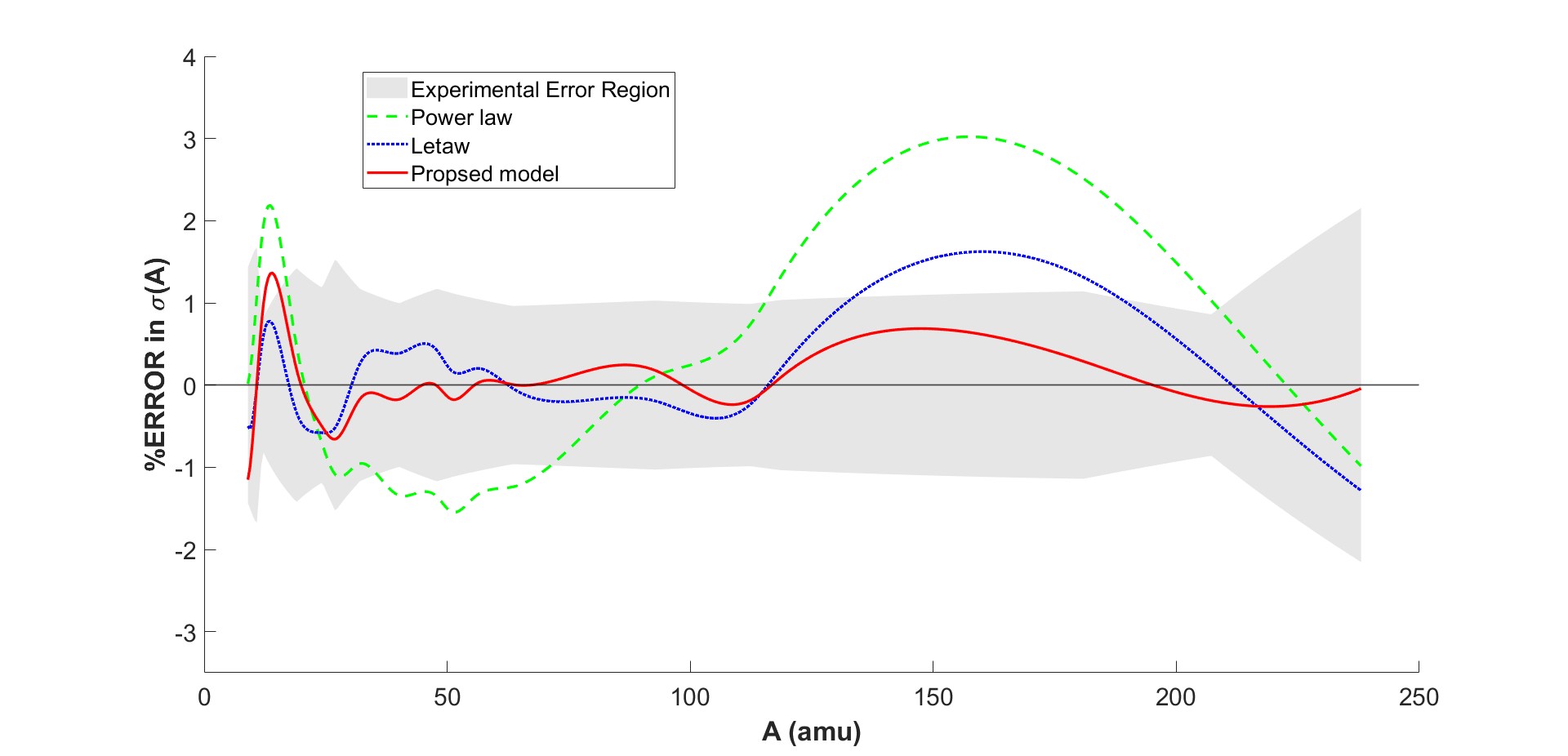}
    \caption{The left figure is high-energy correction Eq.(\ref{eq:highour}) fitting with experimental data\cite{Bobchenko:1979hp}.In the right figure, the green dashed line represents the power-law model, the blue dashed line corresponds to the Letaw $\textit{et. al.}$ model, and the red solid line indicates the proposed model. The shaded gray region denotes the experimental uncertainty band. 
}\label{fig:h_fit}
\end{figure}

Energies above 5~GeV are classified as the high-energy region. The high-energy fit is constrained using proton-nucleus inelastic cross-section data reported by Bobchenko \textit{et al.}~\cite{Bobchenko:1979hp} in the energy range 5-9~GeV. In this regime, the incident proton has a large momentum, and the interaction dynamics are dominated by geometrical considerations rather than detailed nuclear structure effects. Consequently, the inelastic cross-section exhibits a weak dependence on energy and the cross-section has an asymptotic value that depends primarily on the mass number of the target nucleus. The proposed empirical formula for the high-energy region is
\begin{equation}
    \sigma_h(A) = a A^b \left[1 - c \cos(d \sqrt{A}) + e A^f \right],
    \label{eq:highour}
\end{equation}
The functional dependence on the mass number$A$ is parameterized using a combination of power-law and oscillatory terms. The term$ A^b$ represents a geometrical approximation of the cross section and the other term in the bracket is the correction term. The cosine term accounts for the effects of the nuclear structure. The term$ A^f$ becomes relevant primarily for heavier nuclei, and is specifically incorporated to capture the trend of experimental data for heavy elements. 
 The parameters of the model$a, b, c, d, e,$ and$f$ are determined through a least chi-square fitting of the available experimental data for the total inelastic cross section of different atomic masses ($A$) scattered by a proton \cite{Bobchenko:1979hp}. The optimized values of these parameters are listed in Table-\ref{tab:parameters}.  This parameterization achieves a reduced chi-square per degree of freedom is 0.21. 
\begin{table}[h]
\centering
\caption{Model parameters for Eq.(\ref{eq:highour})}
\begin{tabular}{|c|c|c|c|c|c|c|}
\hline
\textbf{Parameter} & $a$ & $b$ & $c$ & $d$ & $e$ & $f$  \\
\hline
\textbf{Value}  & 47.4 & 0.676 & 0.045 & 0.42  & 0.000018  & 1.63\\
\hline
\end{tabular}
\label{tab:parameters}
\end{table}
Fig.\ref{fig:h_fit} represents the least chi-square fitting plot of Eq.(\ref{eq:highour}). The range of cross-section is very long, and an investigation on the percentage variation of the cross-section may shed more light on the precision efficiency.  
Fig.\ref{fig:h_fit} illustrates the percentage error in the proposed model as a function of mass number$ A$  and compares with the Power law model, Letaw $\textit{et. al.}$ model, given by Eq\ref{eq:1} and \ref{eq:2} respectively. The gray band represents experimental error corridor\cite{Bobchenko:1979hp}. The power law model, marked as a green dashed line, shows a very large deviation from the experimental uncertainty band throughout the whole $A$ range. The Letaw $\textit{et. al.}$ model shows good agreement for the lower range $A<130$, but not so consistent for heavier elements. While, the proposed model achieves a consistently lower percentage error across the entire range of elements. 

\subsubsection{Low Energy} 
At low energies(below 2 $GeV$), the total inelastic cross-section in not independent of energy. 
The proton with low momentum increases the probability of interaction with the target nucleus. As a result, the cross-section is higher at low energies.  
All the available experimental measurements for the proton scattered of nucleus of different elements show a larger cross-section at low energies.  Around the range $0.2< E < 2~GeV$, cross-section decreases and shows a minimum of about $15\%$ drop, which eventually saturates for higher energies to the values mentioned in Fig.\ref{fig:h_fit}. At the even lower range below $200~MeV$, the cross-section increases rapidly and shows a sharp peak having more than $60\%$ higher value than the saturation value. Such oscillatory behavior demands precision modeling for the low-energy regime. Based on the interaction scales and region, the complete inelastic cross-section can be factorized as high-energy and low-energy corrections. The proposed empirical formula with low energy correction is 
\begin{equation}
\sigma_{inel}(Z, A, E) = \sigma_h(A).\ X(E).\ Y_{coul}(Z,E)
    \label{eq:loweq}
\end{equation}
where, $ \sigma_h(A)$ is the high-energy contribution as given by Eq.(\ref{eq:highour}), while the function $X(E)$ and $Y_{coul}(Z,E)$ introduce low-energy corrections to the inelastic cross-section. $X(E)$ incorporates two main components: an exponentially decaying term to capture the low-energy behavior, and a $\sin$ term modulated by an exponential decay to maintain flexibility in the fit and reads as
\begin{equation}
X(E) = 1 - \alpha \cdot \exp\left(-\frac{E}{\beta}\right) \cdot \sin\left(\gamma \cdot E^{\lambda} \right) + \frac{1}{2} \cdot \exp\left(-\frac{E}{\epsilon} \right).
\label{eq:X}
\end{equation}
\begin{figure}[h]
    \centering
    \includegraphics[width=0.45\linewidth, trim=100 220 100 220, clip]{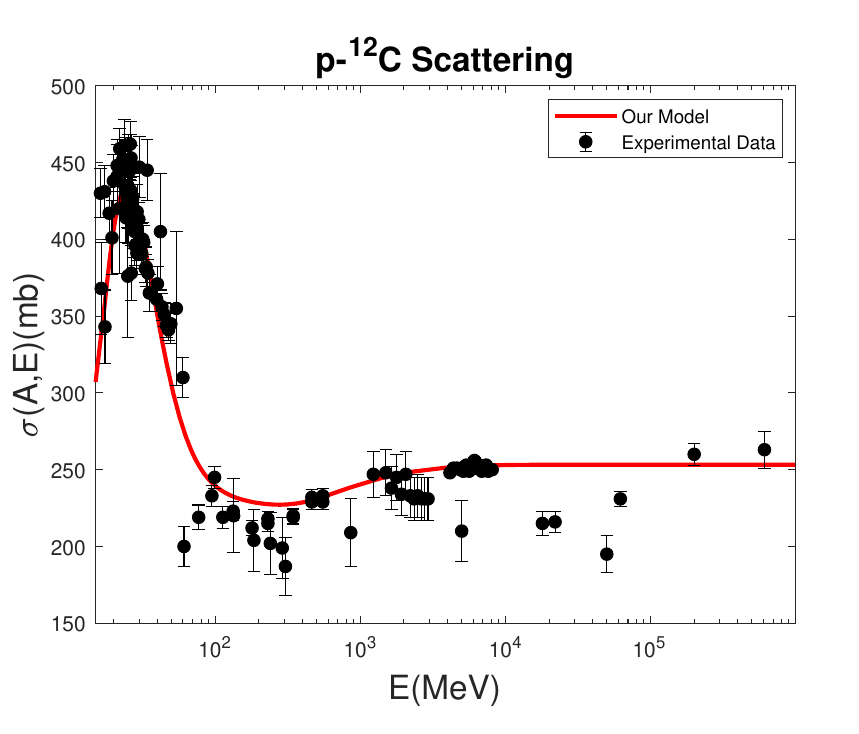}
    \includegraphics[width=0.45\linewidth, trim=100 220 100 220, clip]{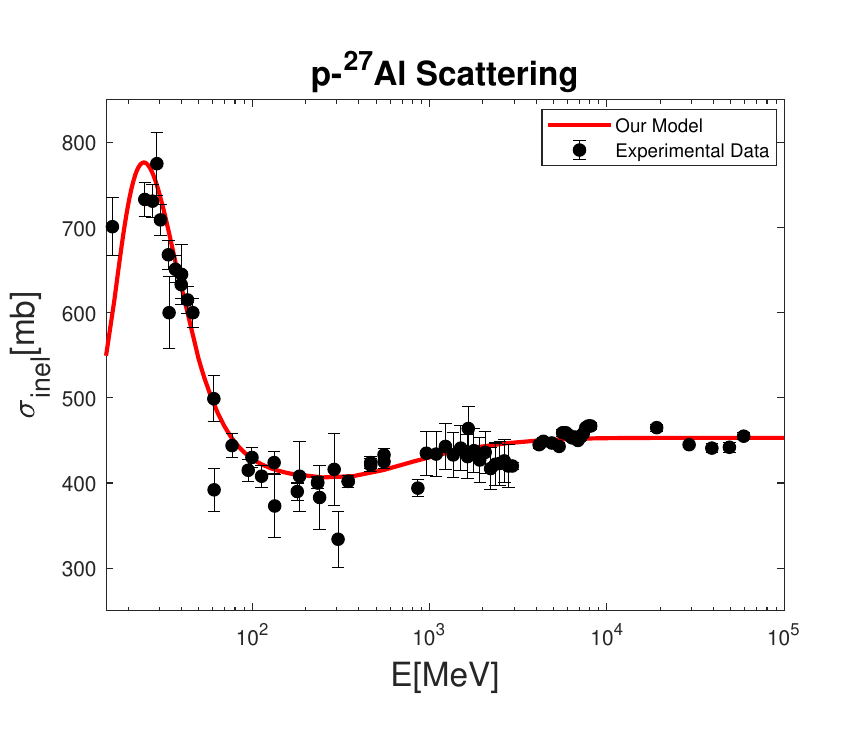}
    \caption{This figure shows the fitting of proton scattering with $^{12}$C and $^{27}$Al experimental data \cite{carlson1996,Bobchenko:1979hp,denisov1973absorption,fumuro1979dependence, Grchurin1985} using the least chi-square method using machine learning.}  
    \label{fig:Al_fit}
\end{figure}
Additionally, the term$Y_{coul}$ is included to care for the Coulomb barrier at low energies for all nuclei and given by
\begin{equation}
Y_{coul}(Z,E) = 1 + k(Z) \cdot \exp{\left(\frac{-E}{24}\right)}
\label{eq:Y}
\end{equation}
The
All parameters$\alpha$,$\beta$,$\gamma$,$\lambda$, and $\epsilon$ are obtained by least chi-square fitting of the experimental data for proton scattering cross-section from Aluminium and Carbon targets using machine learning methods and listed in table-\ref{tab:parameters}. 
Fig.\ref{fig:Al_fit}, represents the fitting of the extensively measured experimental data for Aluminium and Carbon nuclei over a wider range of energy. 
We obtain the minimum chi-square per degree of freedom for carbon and Aluminium are 3.29 and 3.72, respectively.
\begin{table}[h]
\centering
\caption{Model parameters for Eq.(\ref{eq:X})}
\begin{tabular}{|c|c|c|c|c|c|}
\hline
\textbf{Parameter} & $\alpha$ & $\beta$ & $\gamma$ & $\lambda$ & $\epsilon$  \\
\hline
\textbf{Value}  & 0.3 & 3742  & 68  & -0.83 & 134 \\
\hline
\end{tabular}
\label{tab:parameter_lowenergy}
\end{table}
The low-energy oscillatory behavior is also sensitive to the atomic numbers ($Z$) of the target nucleus, which is addressed by the Coulomb term $ Y_{coul}(Z,E)$.  The fine-tuning of the low-energy peak for different elements is controlled by the function $k$, in Eq.(\ref{eq:Y}), as listed in Table-\ref{tab:k_values}.

\section{Numerical Results and Predictions} 
\label{sec:numerical_results} 

In this section, for universality, we discuss our model results and predictions for other target nuclei scattering cross-sections measured experimentally and compare with the existing models. The numerical results are carried out for a broad range of projectile energies and different target nuclei. Special emphasis is given to the energy and mass number dependence of the cross-sections, which are important observables in proton-nucleus interactions.

\begin{figure}[tp]
    \subfigure[]{\includegraphics[width=0.34\linewidth, trim=100 240 100 240, clip]{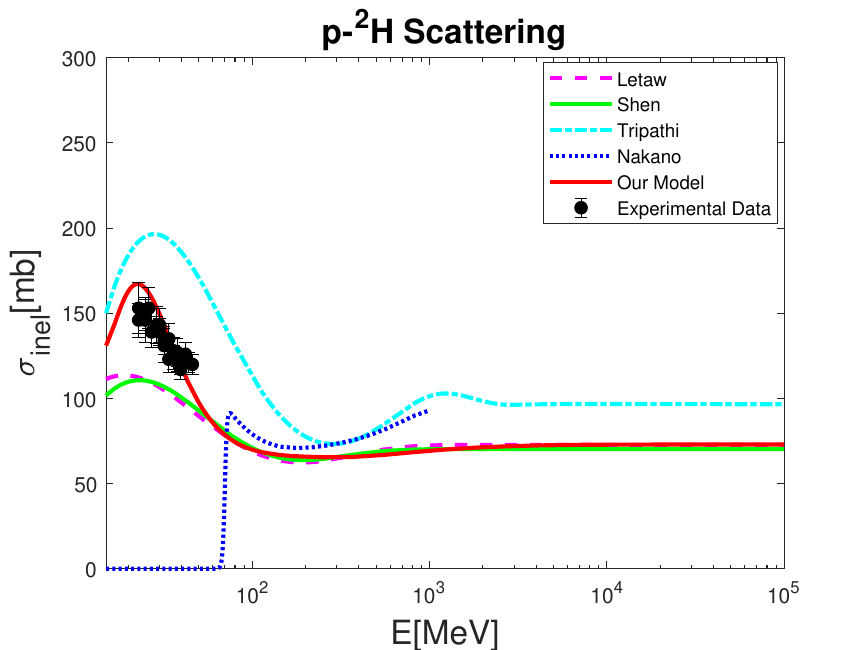}}
    \subfigure[]{\includegraphics[width=0.34\linewidth, trim=100 240 100 240, clip]{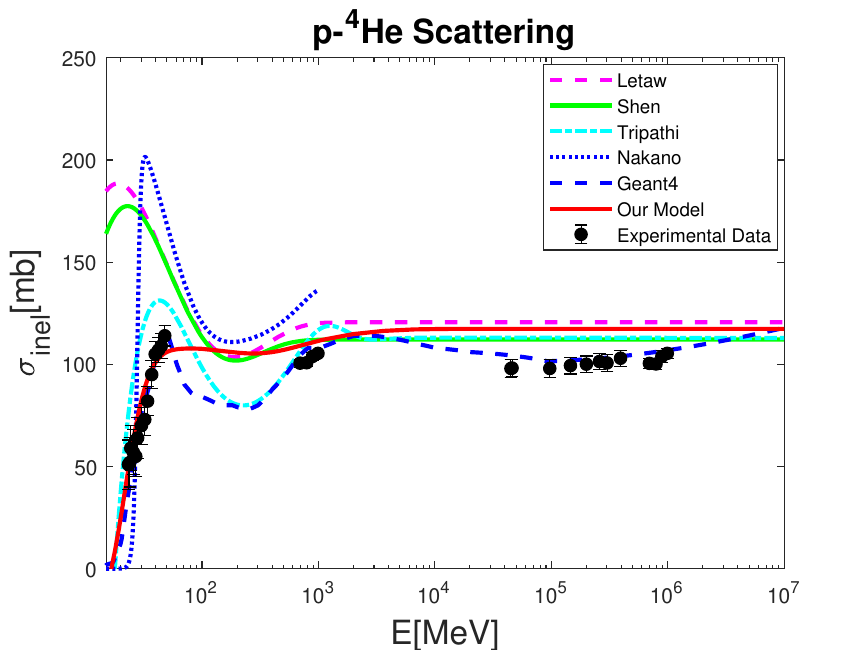}}
    \caption{Model results and comparison for the inelastic scattering cross-section of target (a) $^2$H and (b) $^4$He with configuration $1s$ shell.}
    \label{fig:1s}
\end{figure}

\begin{figure}
\hspace{-1cm}
    \subfigure[]{\includegraphics[width=0.34\linewidth, trim=100 240 100 240, clip]{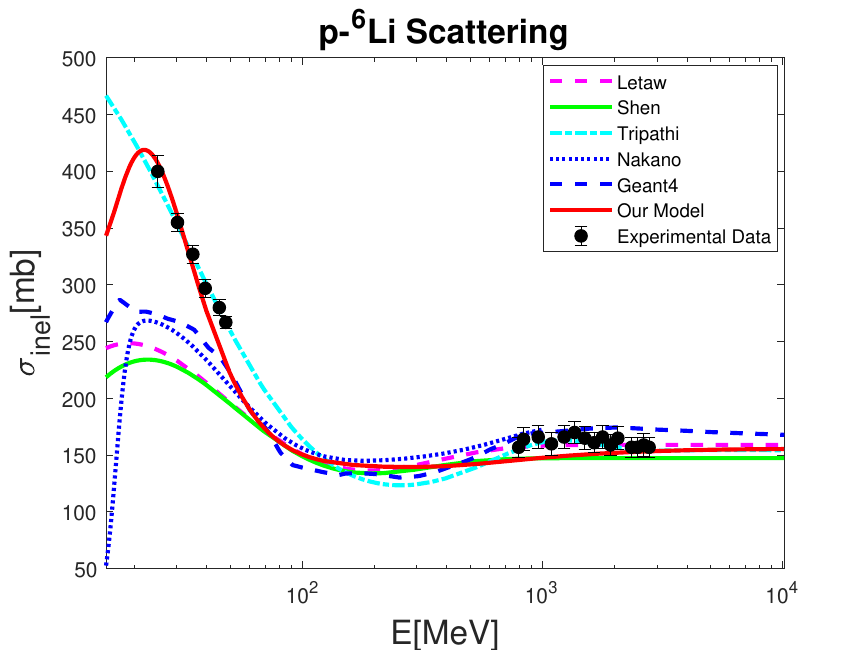}}\hspace{-.5cm}
    \subfigure[]{\includegraphics[width=0.34\linewidth, trim=100 240 100 240, clip]{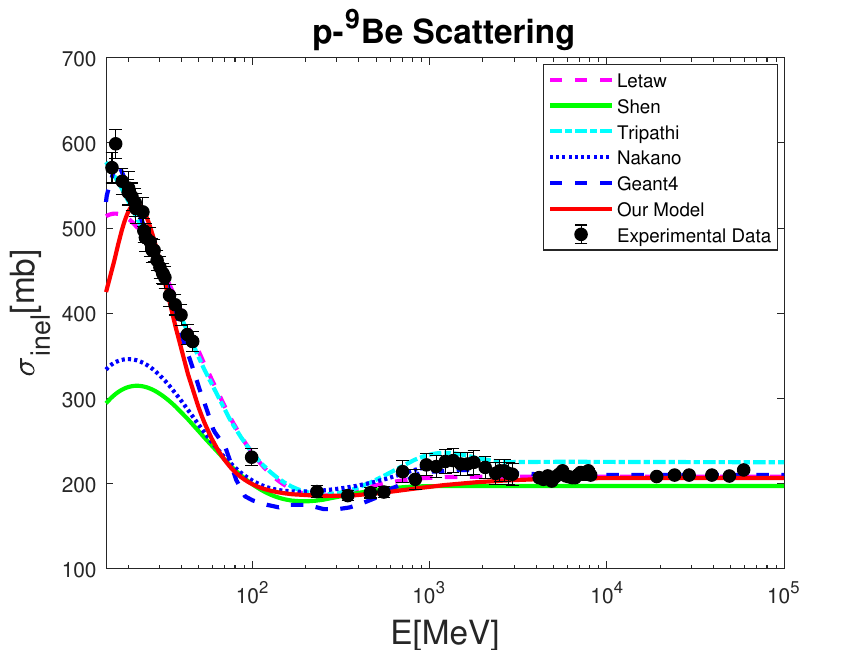}}\hspace{-.5cm}
    \subfigure[]{\includegraphics[width=0.34\linewidth, trim=100 240 100 240, clip]{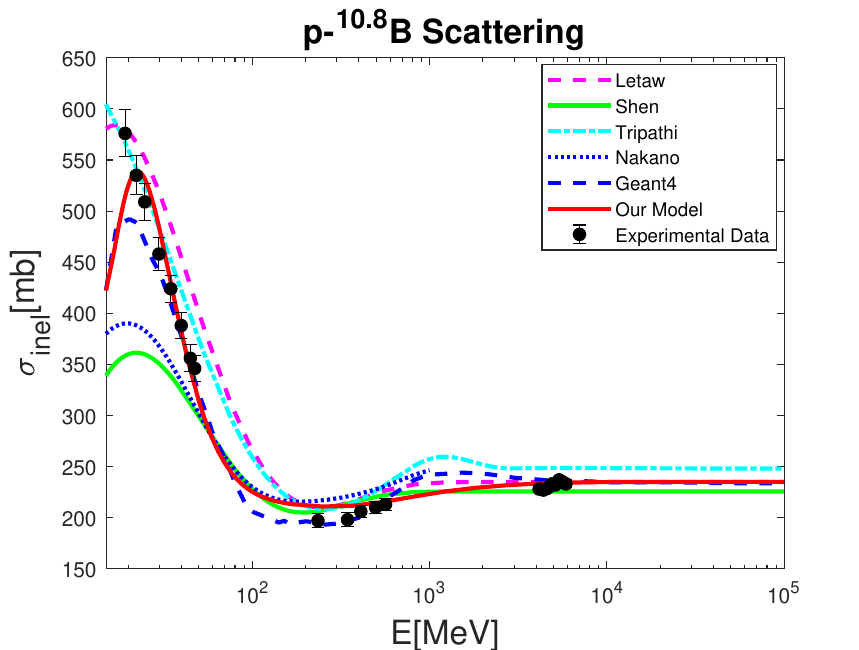}}\\
    \subfigure[]{\includegraphics[width=0.34\linewidth, trim=100 240 100 240, clip]{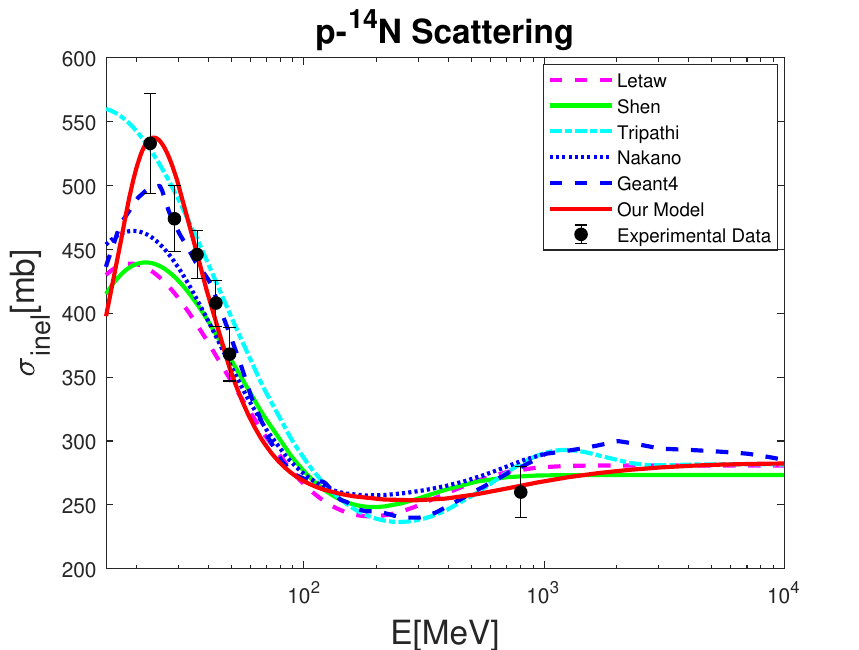}}
    \subfigure[]{\includegraphics[width=0.34\linewidth, trim=100 240 100 240, clip]{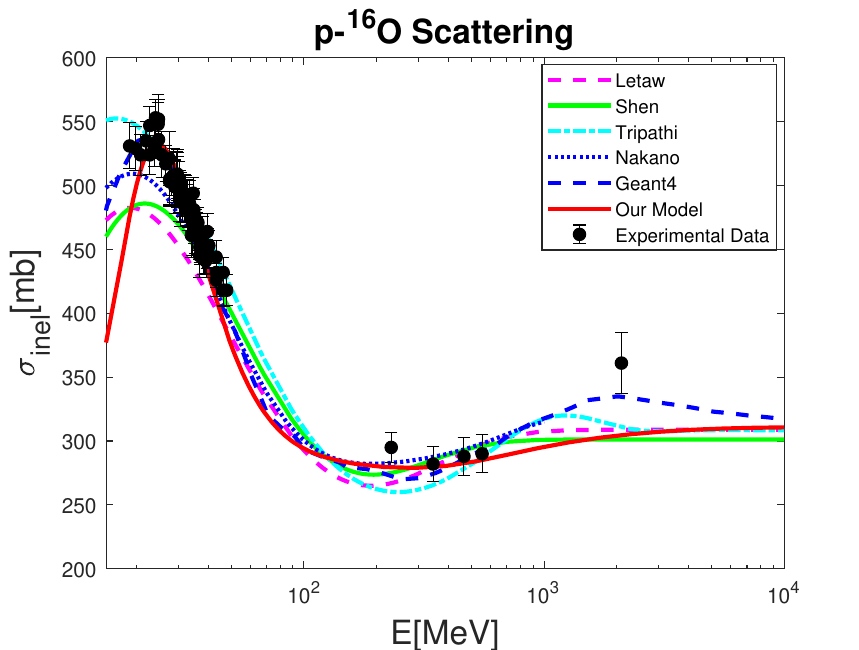}}
    \caption{Model results and comparison for the inelastic scattering cross-section of target (a)$^6$Li, (b)$^9$Be, (c)$^{10.8}$B, (d)$^12$C, (e)$^{14}$N with configuration $1p$ shell.}
    \label{fig:1p}
\end{figure}

\begin{figure}[tp]
\hspace{-1cm}
        \subfigure[]{\includegraphics[width=0.34\linewidth, trim=100 240 100 240, clip]{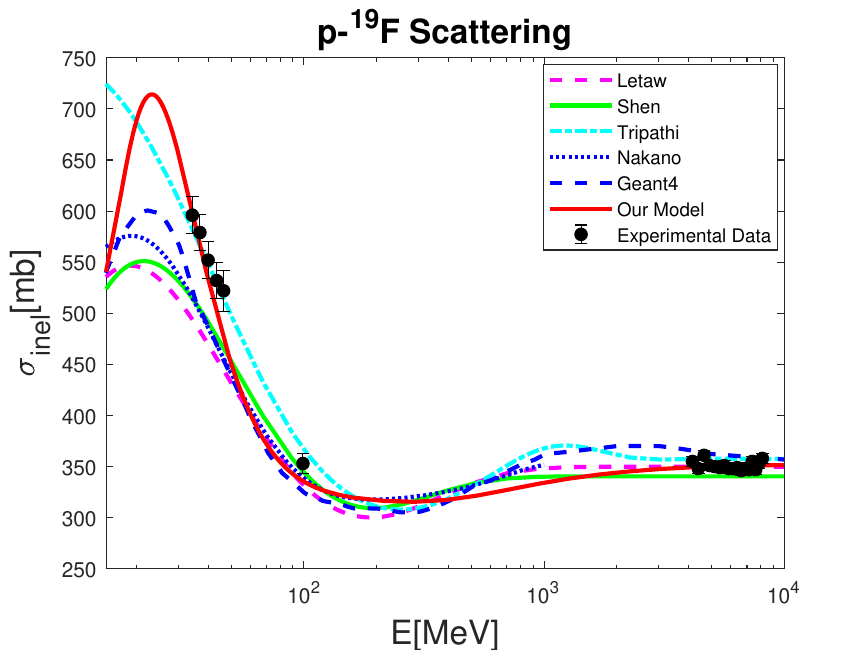}}\hspace{-.5cm}
        \subfigure[]{\includegraphics[width=0.34\linewidth, trim=100 240 100 240, clip]{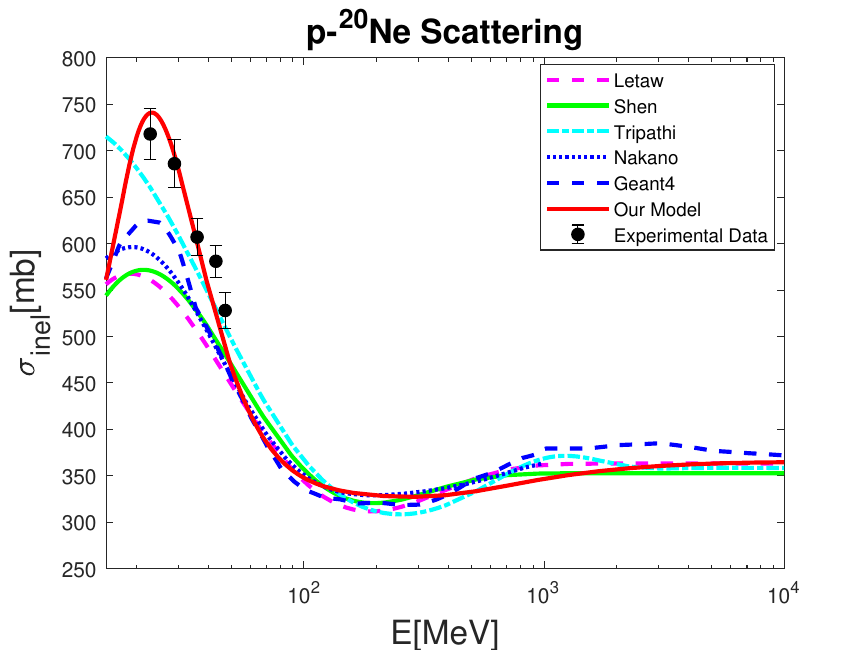}}\hspace{-.5cm}
        \subfigure[]{\includegraphics[width=0.34\linewidth, trim=100 240 100 240, clip]{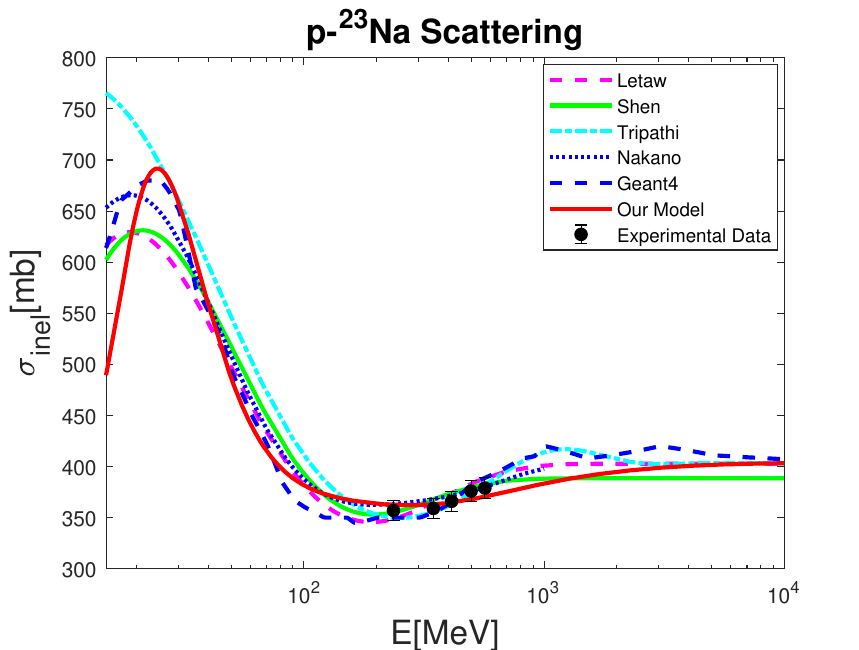}}\\
        \subfigure[]{\includegraphics[width=0.34\linewidth, trim=100 240 100 240, clip]{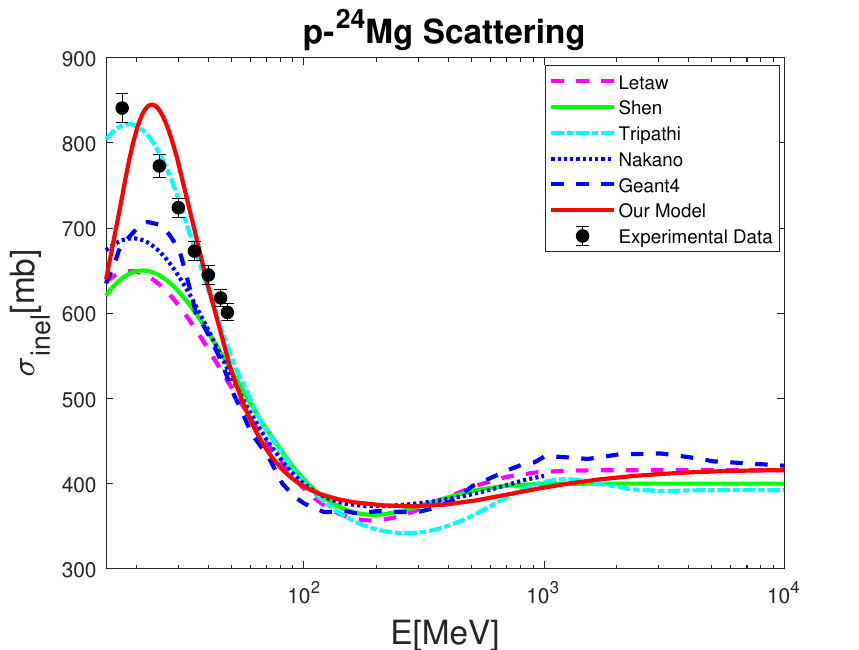}}\hspace{-.5cm}
        \subfigure[]{\includegraphics[width=0.34\linewidth, trim=100 240 100 240, clip]{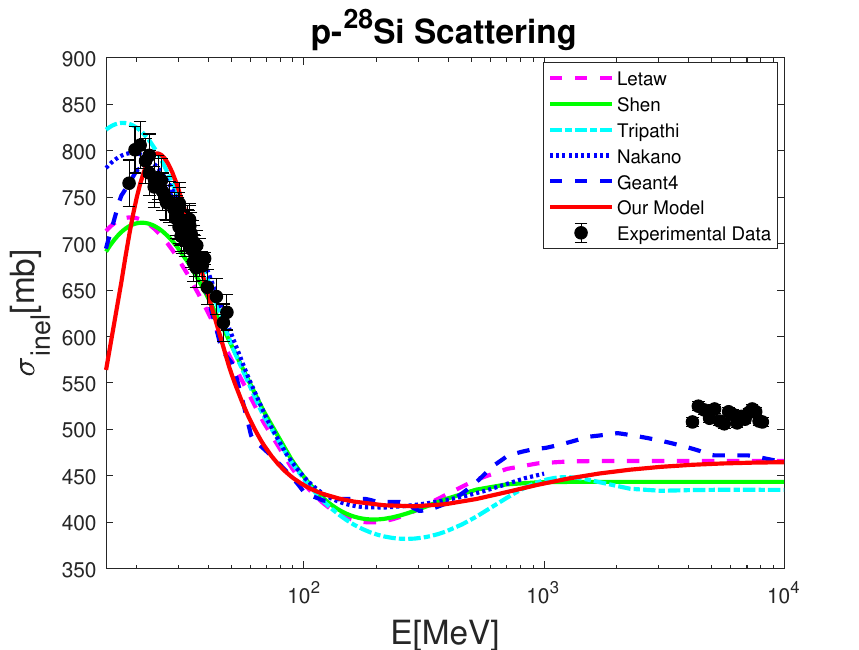}}\hspace{-.5cm}
        \subfigure[]{\includegraphics[width=0.34\linewidth, trim=100 240 100 240, clip]{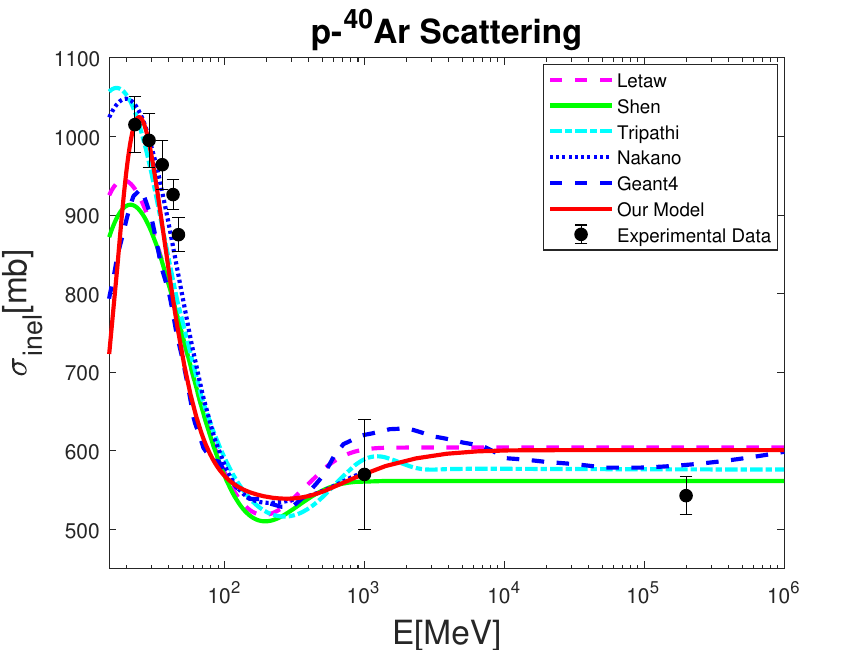}}\hspace{-.5cm}
        \subfigure[]{\includegraphics[width=0.34\linewidth, trim=100 240 100 240, clip]{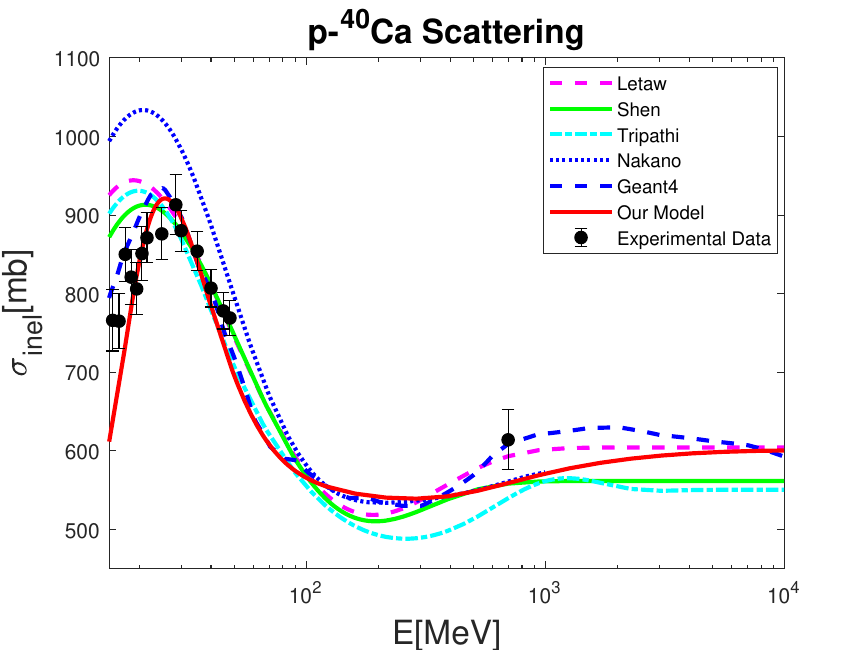}}\\
        \caption{Model results and comparison for the inelastic scattering cross-section of target (a)$^{19}$F, (b)$^{20}$Ne, (c)$^{23}$Na, (d)$^{24}$Mg, (e)$^{28}$Si, (f)$^{40}$Ar, and (g)$^{40}$Ca with configuration $2s+1d$ shell.}
        \label{fig:2s1d}
\end{figure}

\begin{figure}[tp]
        \subfigure[]{\includegraphics[width=0.34\linewidth, trim=100 240 100 240, clip]{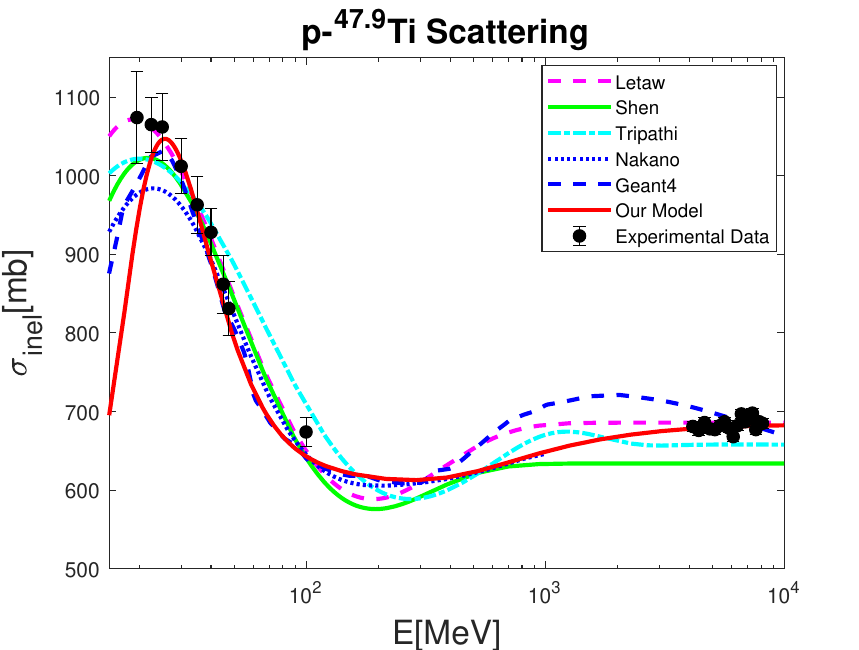}}\hspace{-.5cm}
        \subfigure[]{\includegraphics[width=0.34\linewidth, trim=100 240 100 240, clip]{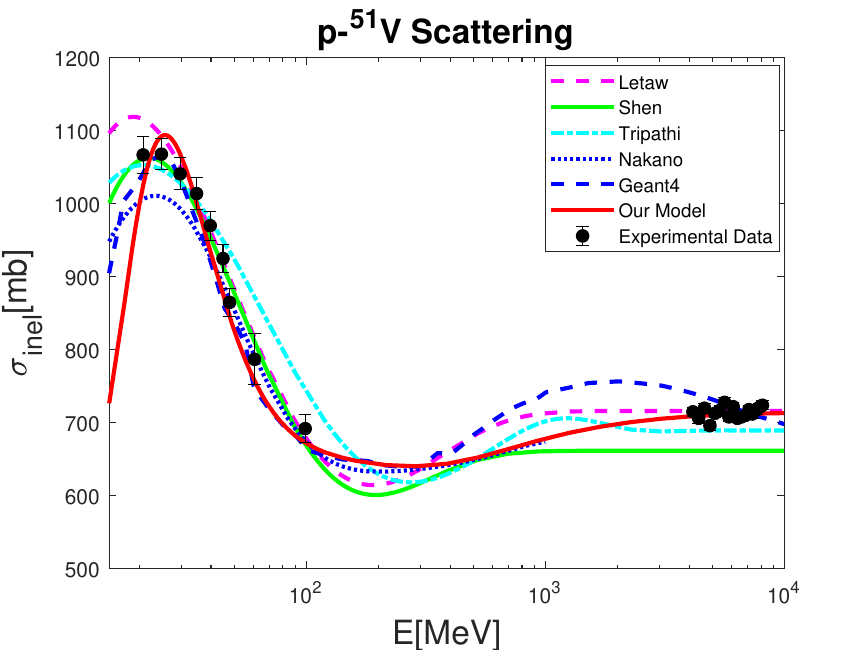}}\hspace{-.5cm}
        \subfigure[]{\includegraphics[width=0.34\linewidth, trim=100 240 100 240, clip]{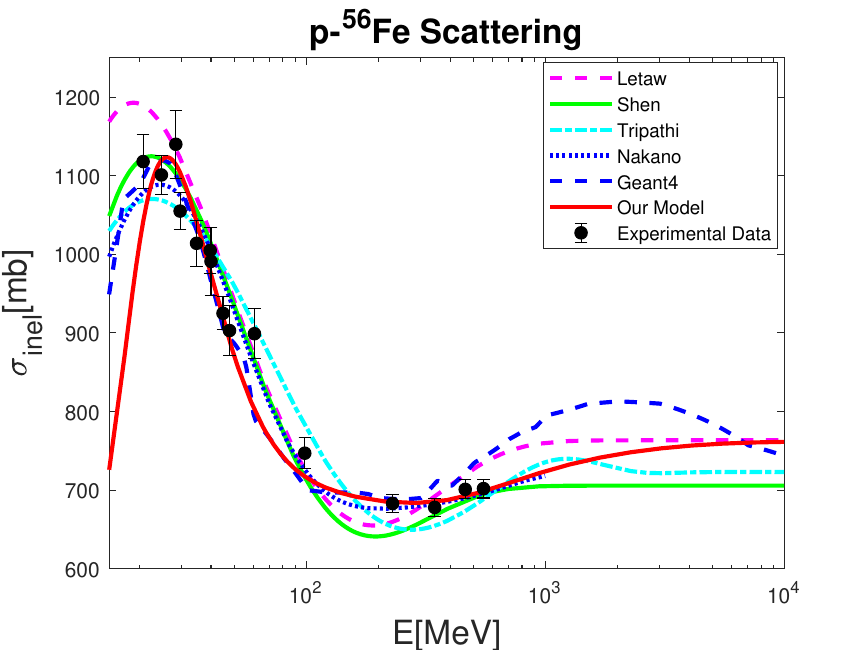}}\\
        \subfigure[]{\includegraphics[width=0.34\linewidth, trim=100 240 100 240, clip]{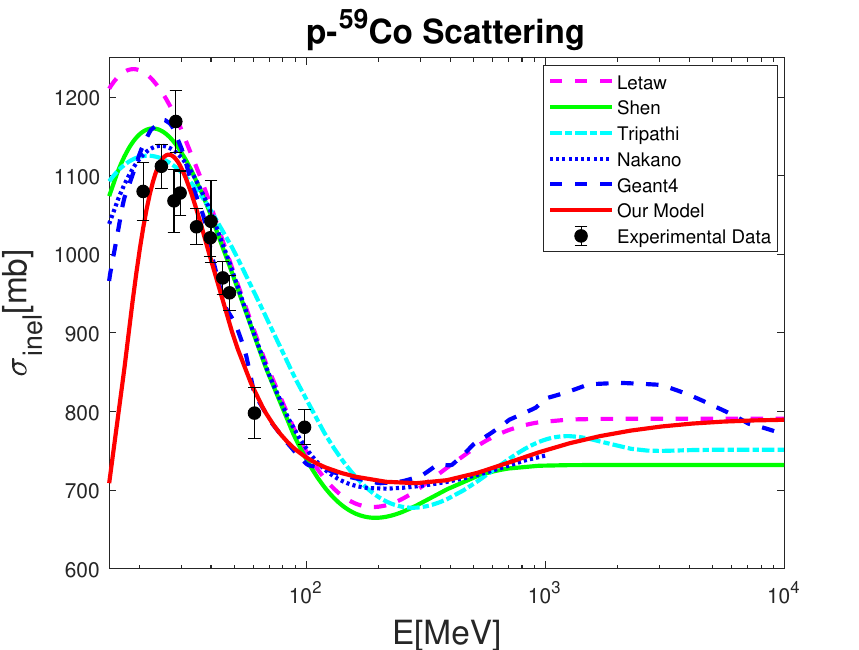}}\hspace{-.5cm}
        \subfigure[]{\includegraphics[width=0.34\linewidth, trim=100 240 100 240, clip]{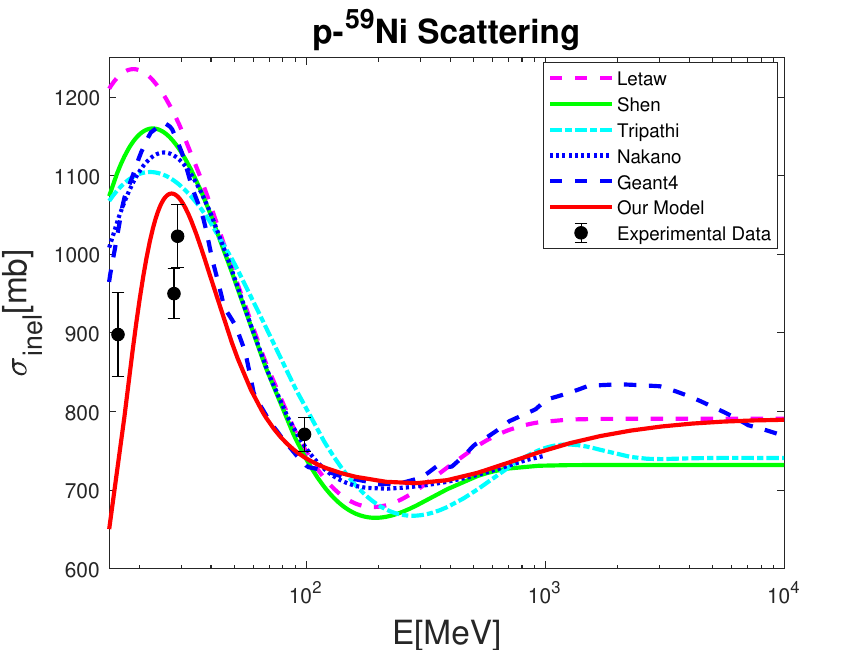}}
        \caption{Model results and comparison for the inelastic scattering cross-section of target (a)$^{47.9}$Ti, (b)$^{51}$V, (c)$^{56}$Fe, (d)$^{59}$Co, and (e)$^{59}$Ni with configuration $1f_{7/2}$ shell.}
        \label{fig:1f7/2}
\end{figure}

\begin{figure}[tp]       
        \subfigure[]{\includegraphics[width=0.34\linewidth, trim=100 240 100 240, clip]{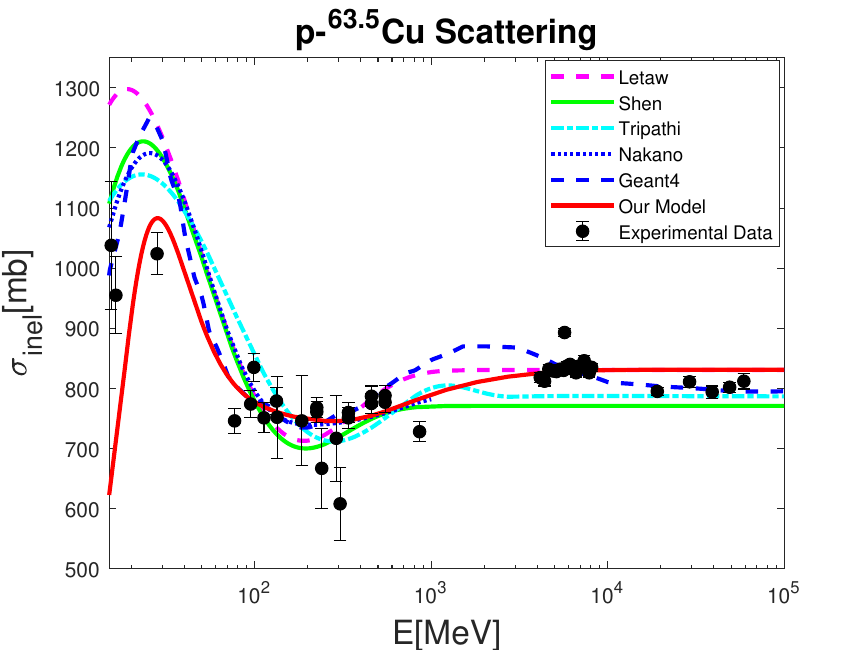}}\hspace{-.5cm}
        \subfigure[]{\includegraphics[width=0.34\linewidth, trim=100 240 100 240, clip]{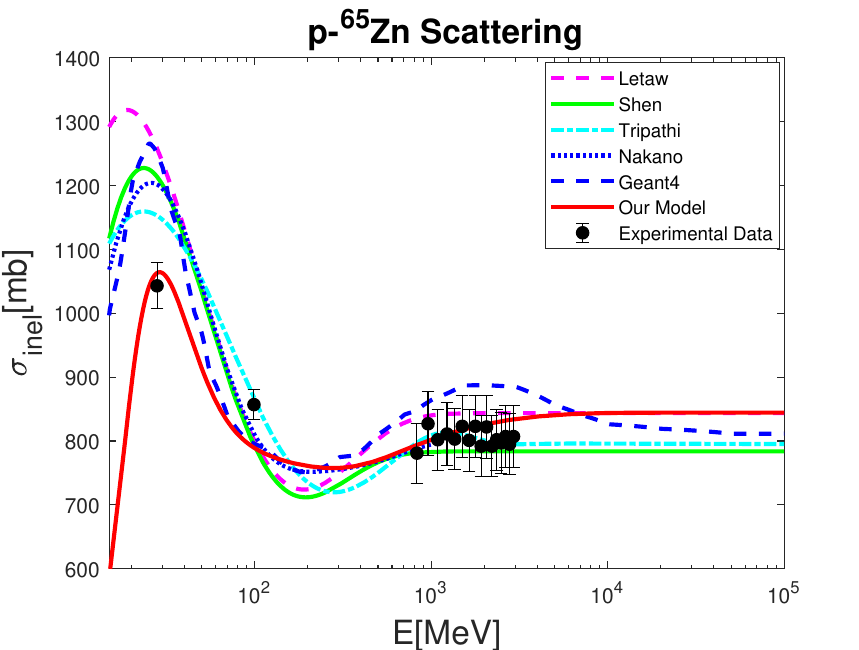}}\hspace{-.5cm}
        \subfigure[]{\includegraphics[width=0.34\linewidth, trim=100 240 100 240, clip]{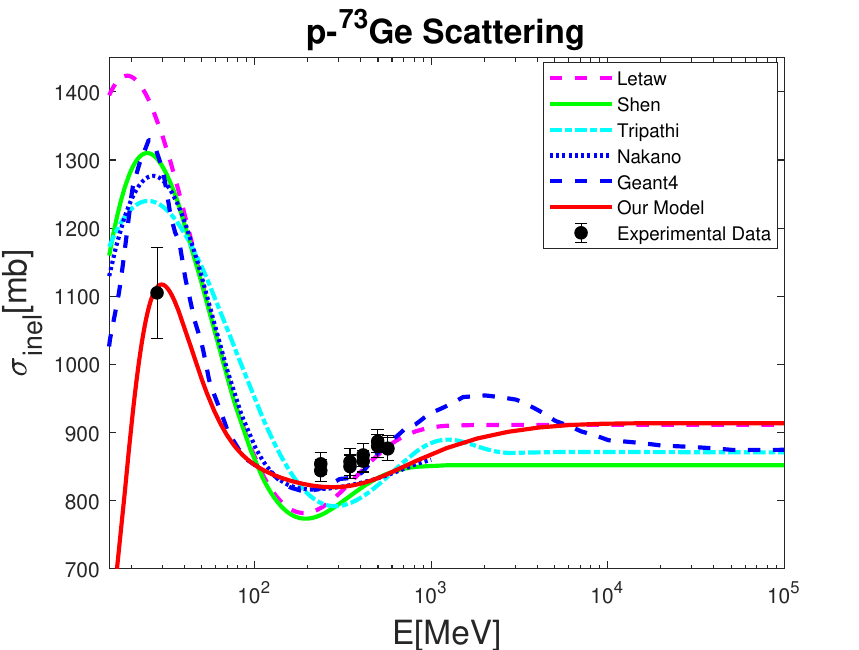}}
        \caption{Model results and comparison for the inelastic scattering cross-section of target (a)$^{63.5}$Cu, (b)$^{65}$Zn, and (c)$^{73}$Ge with configuration $2p+1f_{5/2}$ shell.}
        \label{fig:2p1f5/2}
\end{figure}

\begin{figure}[tp]
        \subfigure[]{\includegraphics[width=0.34\linewidth, trim=100 240 100 240, clip]{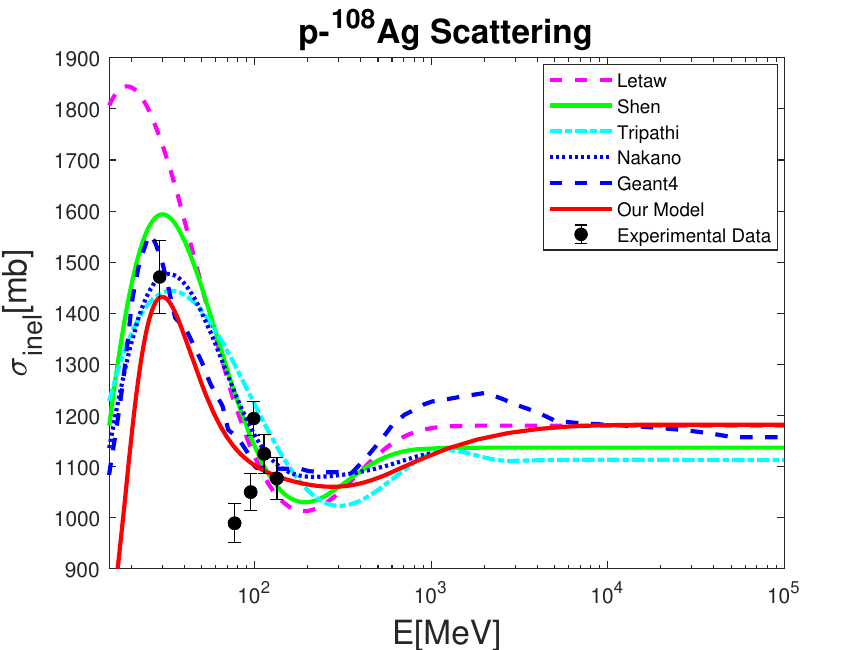}}\hspace{-.5cm}
        \subfigure[]{\includegraphics[width=0.34\linewidth, trim=100 240 100 240, clip]{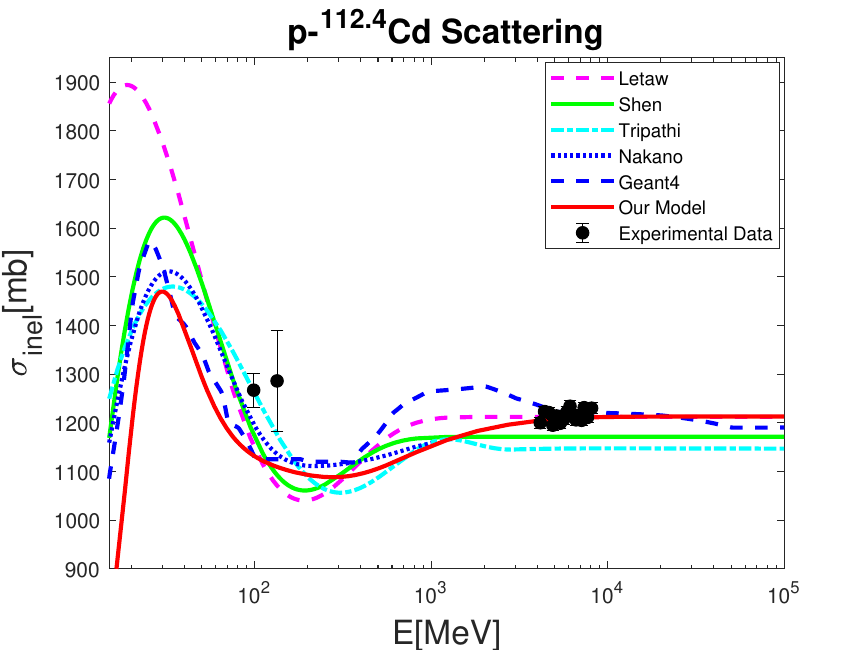}}\hspace{-.5cm}
        \subfigure[]{\includegraphics[width=0.34\linewidth, trim=100 240 100 240, clip]{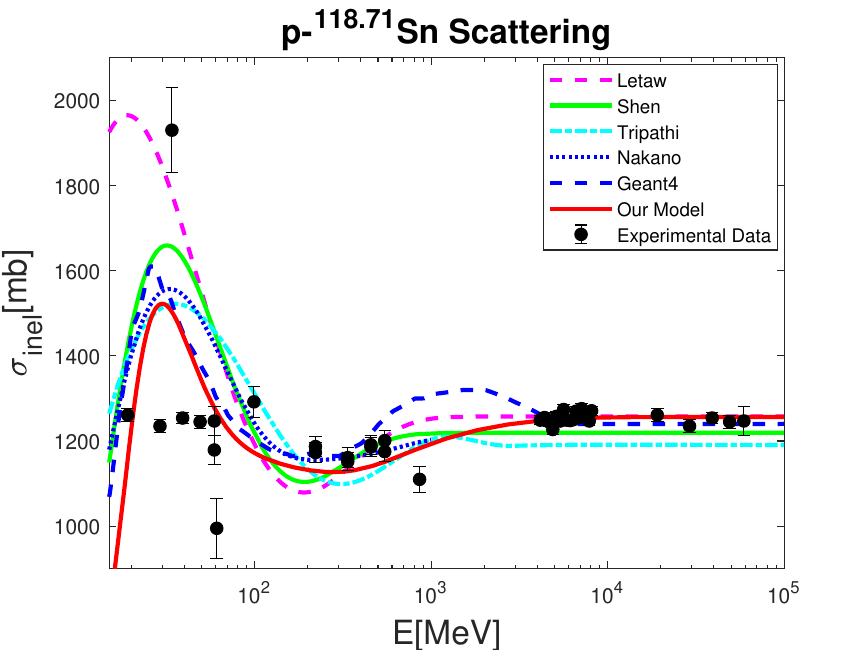}}
        \caption{Model results and comparison for the inelastic scattering cross-section of target (a)$^{108}$Ag, (b)$^{112.4}$Cd, and (c)$^{118.7}$Sn with configuration  $1g_{9/2}$ shell.}
        \label{fig:1g9/2}
\end{figure}

\begin{figure}
        \subfigure[]{\includegraphics[width=0.34\linewidth, trim=100 240 100 240, clip]{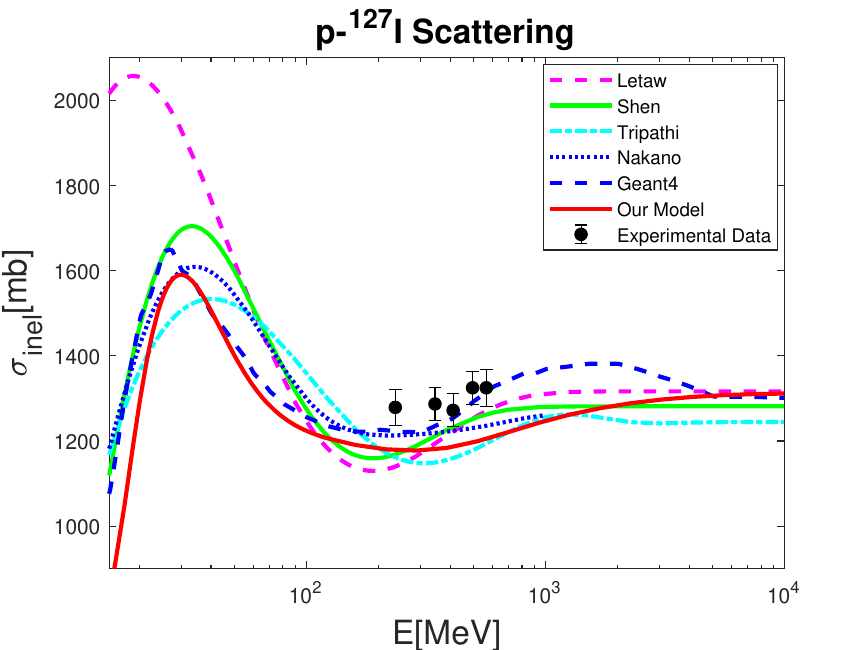}}\hspace{-.5cm}
        \subfigure[]{\includegraphics[width=0.34\linewidth, trim=100 240 100 240, clip]{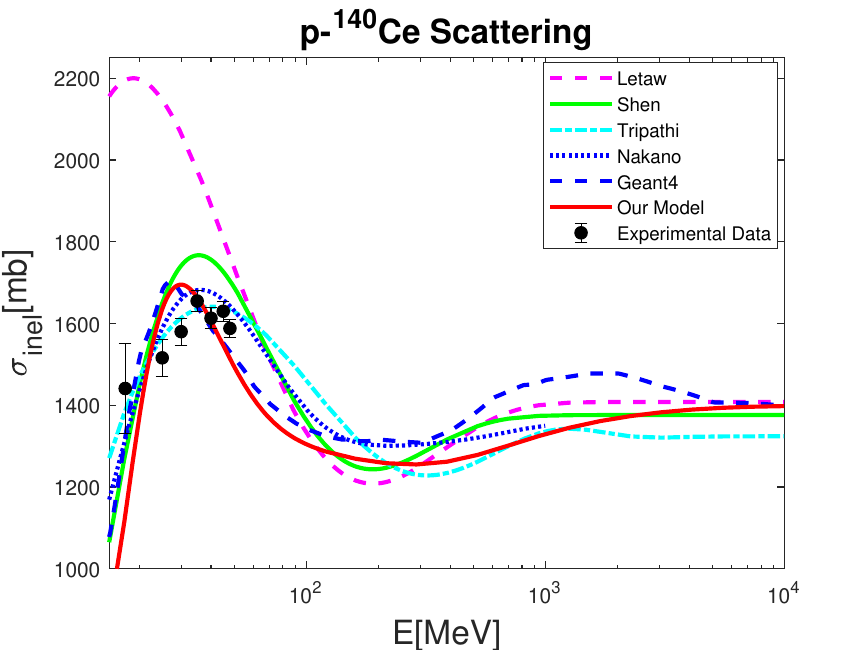}}\hspace{-.5cm}
        \subfigure[]{\includegraphics[width=0.34\linewidth, trim=100 240 100 240, clip]{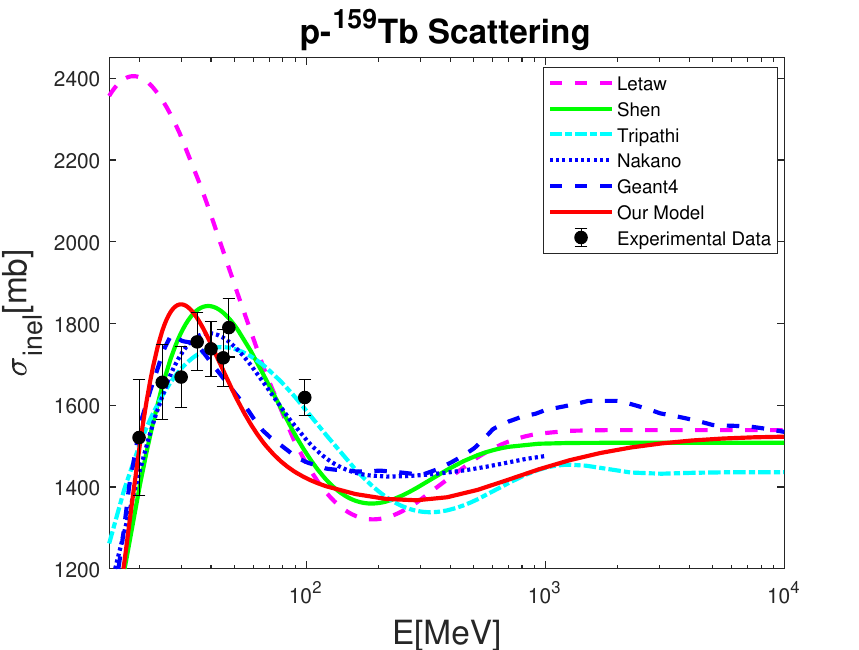}}\\
        \subfigure[]{\includegraphics[width=0.34\linewidth, trim=100 240 100 240, clip]{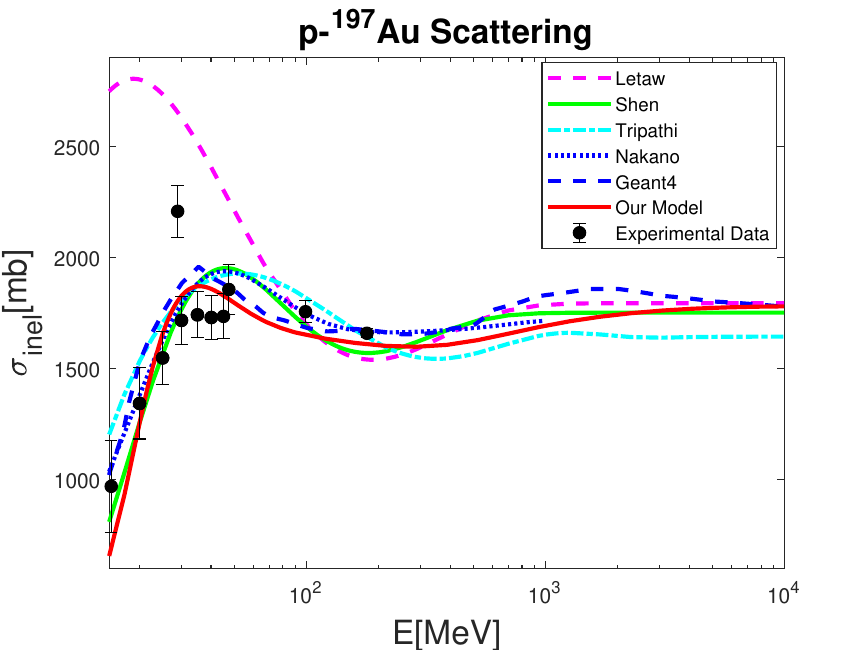}}\hspace{-.5cm}
        \subfigure[]{\includegraphics[width=0.34\linewidth, trim=100 240 100 240, clip]{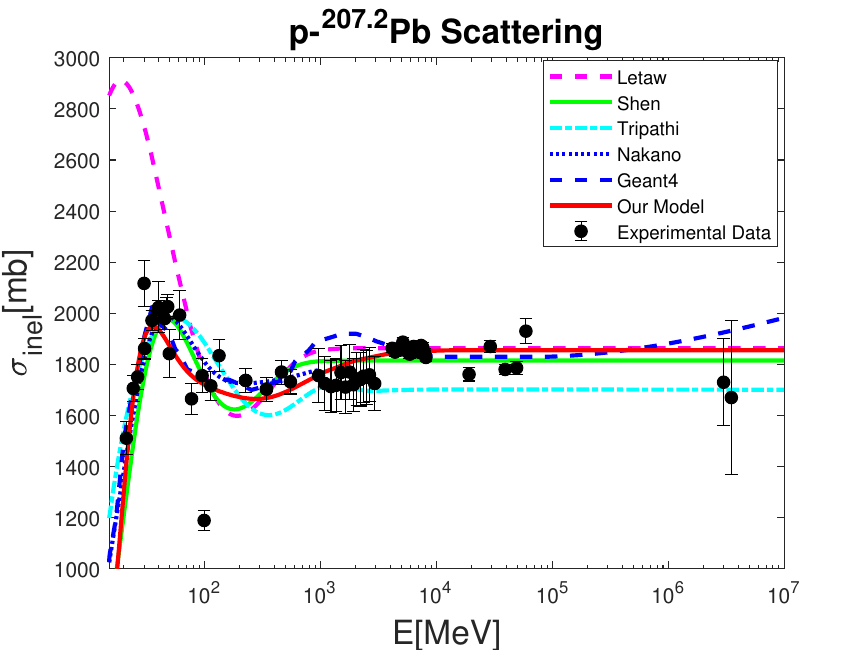}}
        \caption{Model results and comparison for the inelastic scattering cross-section of target (a)$^{127}$I, (b)$^{140}$Ce, (c)$^{159}$Tb, (d)$^{197}$Au, and (e)$^{207.2}$Pb with configuration $3s+2d+1g_{7/2}+1h_{11/2}$ shell.}
         \label{fig:h11/2}
\end{figure}

\begin{figure}[h]
    \subfigure[]{\includegraphics[width=0.35\linewidth, trim=100 240 100 240, clip]{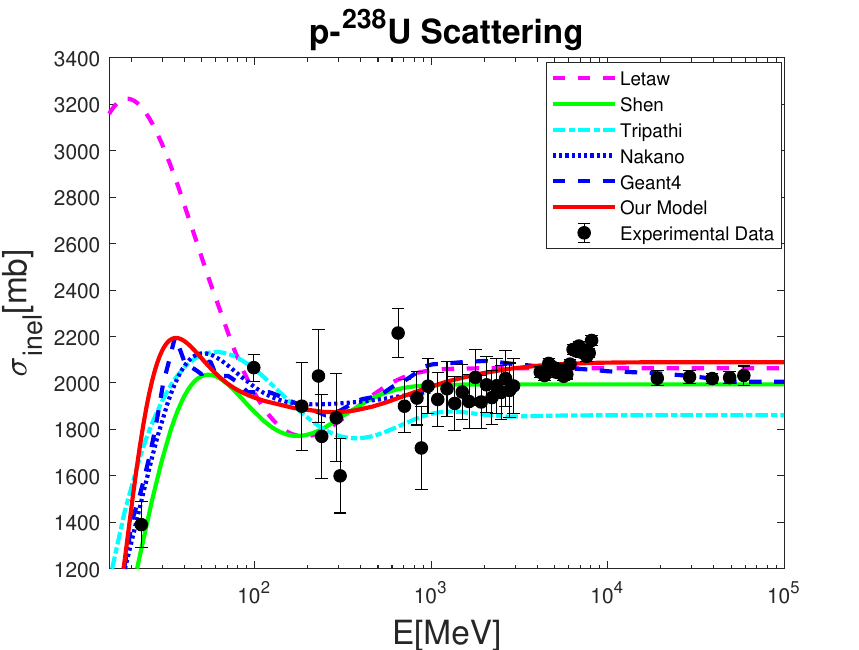}}
    \caption{Model results and comparison for the inelastic scattering cross-section of target (a)$^{238}$U with configuration $1h_{9/2}+2f_{7/2}+1i_{13/2}$ shell.}
    \label{fig:i13/2}
\end{figure}

In Figs.\ref{fig:1s}-\ref{fig:i13/2} represent variation of $p-N$ inelastic cross-sections for different elements of mass A marked as p-$^A$N. 
The x-axes represent the kinetic energy of the incident proton (in MeV), typically on a logarithmic scale to accommodate the broad energy range and to better visualize low-energy and high-energy regions simultaneously. The y-axis shows the inelastic cross-section ($\sigma_{inel}$ in millibarns), which is the observable of interest. The black dots are the experimental data, with the error bar representing the error in cross-section at a particular energy. In the plot, multiple curves are shown corresponding to different models: the pink dashed line shows Letaw $\textit{et. al.}$’s empirical formula, the green solid line shows Shen’s model, the sky blue dot-dashed line shows Tripathi $\textit{et. al.}$’s formulation, the blue dotted line shows Nakano $\textit{et. al.}$’s correction-based model, the blue dashed line shows GEANT4’s simulation, and the red solid line shows our proposed model.

The experimental data used in this work were collected from different sources mentioned in Table-\ref{table: experimental data citations}, as listed in the Appendix. The accuracy of the experimental data of the cross sections depends on several factors, such as target (e.g., thickness), energy spread of the incident beam, and background subtraction. For example, the $^{2}$H data employed here were measured using CD$_2$ targets, since pure hydrogen targets are difficult to prepare with sufficient density and thickness, which leads to challenges in highly precise measurements. The experimental data obtained from \cite{Bobchenko:1979hp} have been employed for most of the elements (C, Pb, Al, B, Be, F, Si, U, Cd, Cu, V, Ti, Sn) in the high–energy region, where the measurements are highly precise. However, in the low–energy region (around 20 MeV), the data available for Au \cite{abegg1979} are comparatively less accurate. The uncertainties in the experimental data are represented by error bars in the plots: larger error bars correspond to less precise data, whereas smaller error bars indicate higher precision.
In the low-energy region, elements such as Mg, He, Sn, and Ce show relatively small uncertainties, while other nuclei exhibit large experimental errors. At higher energies, Zn data are particularly imprecise that is taken from Ref.\cite{Grchurin1985}, and for most elements, the available high-energy data are either low or entirely missing. Certain elements— including  N, Ne, Na, Mg, Ar, Ni, Ge, Ag, I, Ce, and Tb- have limited data coverage at both high- and low-energy regions. These limitations in the experimental database affect the model precision.

The numerical results of this model and comparison, for all these elements, are presented according to the nuclear periodic table that reflects the final element as the magic numbers\cite{Hagino2020NuclearPeriodicTable, PhysRevC.100.044315}. The rows of the nuclear periodic table are constructed based on the number of protons in the final state with configuration $1s$, $1p$, $2s+1d$, $1f_{7/2}$, $2p+1f_{5/2}$, $1g_{9/2}$, $3s+2d+1g_{7/2}+1h_{11/2}$, and $1h_{9/2}+2f_{7/2}+1i_{13/2}$ \cite{Hagino2020NuclearPeriodicTable, PhysRevC.100.044315}. Where the unspecified levels without $j$, stand for the combination of the states with all possible $j$, e.g. $d$ for $d_{5/2}+d_{3/2}$ and so on.
In the first row of the nuclear periodic table, the nuclei exhibit a $1s$ configuration, for which only two bound systems exist, namely $^2$H and $^4$He. For these nuclei, our model results of the $ p-N$ scattering cross-section and a comparative study with available models are shown in Fig.~\ref{fig:1s}. All the color codes and symbols indicate the same as mentioned in the previous discussion. 
The left plot of Fig.~\ref{fig:1s} represents p–$^2$H scattering cross-section. Experimental data are available only at low incident proton energies ($E \lesssim 100$~MeV) from Ref.~\cite{carlson1996}. In this energy range, the Letaw $\textit{et. al.}$ and Shen models systematically underestimate the experimental cross sections, while the Tripathi $\textit{et. al.}$ model overestimates the data. The Nakano $\textit{et. al.}$ model predicts vanishing cross sections below approximately 90~MeV, leading to a clear disagreement with the experimental observations at low energies. Furthermore, the Nakano $\textit{et. al.}$ model provides predictions up to about 1~GeV, beyond which, the cross section diverges and becomes unphysical. In contrast, our model reproduces the experimental data well in the low-energy region and exhibits a saturation behavior ($\sim 72.9~mb$) at higher energies that is consistent with the Letaw $\textit{et. al.}$ and Shen models. GEANT4 results are not available for deuteron but are available for other elements.
The right plot (of Fig.~\ref{fig:1s}) is for cross-section of the first magic number Helium ($^4$He) target, which is a stable nucleus. The experimental data are available over a wide energy range extending up to 1~TeV. The Letaw $\textit{et. al.}$, Shen, and Nakano $\textit{et. al.}$ models display pronounced peaks that significantly deviate from the experimental data across most of the energy range. The Tripathi $\textit{et. al.}$ model shows reasonable agreement only in the vicinity of $E \sim 1$~GeV but fails to reproduce the data at both lower and higher energies.
Our model shows reasonably good agreement with the data below 1~GeV; however, at higher energies it exhibits a deviation of approximately $17\%$ from the experimental measurements. While our GEANT4 prediction shows an excellent description of the experimental cross sections over the entire energy range considered. A possible reason would be that the used GEANT4 simulation toolkit  \texttt{G4BGGNucleonInelasticXS} combines two different descriptions for low energy (below approximately $100\,\mathrm{GeV}$) and higher energy regime \cite{Barashenkov1994,Grichine2009}. 

The second row of the nuclear periodic table corresponds to nuclei with valence nucleons occupying the $1p$ shell. The six occupancy is filled by the six elements $^6$Li, $^9$Be, $^{10.8}$B, $^12$C, $^{14}$N, and the magic nucleus $^{16}$O. The empirical formula parameters are determined by the fitting of p-$^12$C scattering data and excluded from this table of results.  Our model results and comparison for the remaining five target nucleus cross-sections are presented in Fig.~\ref{fig:1p}.

The p-$^6$Li system represents a light and weakly bound nucleus 
As shown in Fig.~\ref{fig:1p}-(a), experimental data are available mainly at low energies below 100~MeV and around a few GeV. In the high-energy region, all the models and GEANT4 predictions exhibit similar trends. However, at low-energy this empirical formula and Letaw $\textit{et. al.}$ provide relatively good agreements with the experimental data.
For p-$^9$Be scattering, which involves a loosely bound nucleus with a pronounced $\alpha$-cluster substructure. The p-$^9$Be scattering is extensively studied in several experiments, and the most precise measurement, with uncertainties below 1\%, is performed by Refs.~\cite{Bobchenko:1979hp,Slaus:1975zz}. The Shen and Nakano $\textit{et. al.}$ models underestimate the cross sections in the energy range below 100~MeV. The remaining models show better agreement in this region. particularly, the GEANT4 and Tripathi $\textit{et. al.}$ models reproduce the peak values more accurately. Our model predicts a peak around 25~MeV with a maximum cross section of approximately 525~mb, while the experimental data indicate a higher peak value of about 600~mb at 15~MeV.  At energies of a few GeV, a secondary significant peak is observed in the experimental data, which is reasonably described by this GEANT4 simulation. With the inclusion of a correction term, the Letaw $\textit{et. al.}$ model also provides a satisfactory description over the entire energy range, although it does not reproduce the first peak around 20~MeV accurately\cite{letaw1983}. 
The p-$^{10.8}$B system represents a transition from very light nuclei to more stable $p$-shell nuclei. As seen in Fig.~\ref{fig:1p}, the amount of experimental data is limited, and the measured cross sections follow trends similar to those observed for lighter systems. Notably, in the energy region below 100~MeV, the experimental data tend to align more closely with the predictions of our model than other models.
For the target $^{14}$N, a relatively well-bound $p$-shell nucleus, the available experimental data are sparse and carry relatively large uncertainties \cite{carlson1996}. As illustrated in Fig.~\ref{fig:1p}-(d), the Letaw $\textit{et. al.}$, Shen, and Nakano $\textit{et. al.}$ models predict lower cross-section values near the peak region. In contrast, the Tripathi $\textit{et. al.}$ and GEANT4 models yield results that are closer to the experimental data. The predictions of our model lie well within the experimental error bars. At higher energies, only a single data point is available around 800~MeV, which is reasonably reproduced by almost all the models considered.
The p-$^{16}$O system is of particular importance due to its closed-shell structure. A large number of experimental data points are available at energies below 60~MeV, providing a stringent test of low-energy model predictions. Fig.~\ref{fig:1p}-(e), the Letaw $\textit{et. al.}$, Shen, and Nakano $\textit{et. al.}$ models fail to reproduce the low-energy data. In contrast, the Tripathi $\textit{et. al.}$, GEANT4, and our model show good agreement with the experimental measurements in this region. Overall, compared to other model predictions,  our empirical model and GEANT4 predictions with FTFP$\_$BERT physics list provide a consistent description for $1p$ cell elements of the nuclear periodic table across the energy range. 

The next row of the nuclear periodic table corresponds to nuclei with valence nucleons occupying the $2s$ and $1d$ shells. In this row, a total of twelve elements occupy this state with closed shells at the magic-number nuclei $^{40}$Ca. We present our model result for the targets $^{19}$F, $^{20}$Ne, $^{23}$Na, $^{24}$Mg, $^{28}$Si, $^{40}$Ar, and $^{40}$Ca whose proton scattering cross-section are experimentally measured.
For target $^{19}$F, experimental data are available in two distinct energy regions, namely from 30 to 100~MeV and the saturation region of range 3 to 10~GeV, as shown in sub-Fig.~\ref{fig:2s1d}-(a). In the low-energy region, only the Tripathi $\textit{et. al.}$ model and our model reproduce the experimental data reasonably well, while the remaining models underestimate. The $^{20}$Ne and $^{24}$Mg targets, proton scattering cross-section is measured for low-energy region only, and experimental data are extremely limited in the energy range from 20 to 60~MeV Fig.~\ref{fig:2s1d}-(b,d).  For both cases, our model exhibits smaller deviations from the experimental data compared to the other approaches considered.

Whereas, p-$^{23}$Na scattering cross-section is measured for the moderate energy region between 200 and 600~MeV as seen in Fig.~\ref{fig:2s1d}-(c). All the models provide comparable predictions and reproduce the experimental data reasonably well within this limited energy range. 
However, due to the lack of experimental data, it is not possible to draw definitive conclusions regarding the overall performance of the models across the full energy spectrum and demands further experimental measurement in the future. 
For p-$^{28}$Si scattering cross-section, shown in Fig.~\ref{fig:2s1d}, a relatively large amount of experimental data is available at low energies below 100~MeV and at high energies in the range from 3 to 10~GeV, while no data exist in the intermediate-energy region. Except Letaw $\textit{et. al.}$ and Shen models, other results show a consistent agreement with the low-energy data.  However, none of the models successfully reproduces the saturation value at the high-energy compare to the experimental data. The saturation value for $^{28}$Si differs by about 19$\%$ to the model results. 
In the case of p-$^{40}$Ar scattering cross-section, the available experimental data cover a broad energy range but have a relatively low number of data points with large uncertainties \cite{L} as shown in Fig.~\ref{fig:2s1d}-(d). Nakano $\textit{et. al.}$ model provides the best overall agreement with the experimental data, however the Tripathi $\textit{et. al.}$ model and our model yield predictions that are close to the measurements but do not fully reproduce them within the experimental uncertainties.
The final nucleus in this configuration is $^{40}$Ca, which is a doubly magic nucleus. For p-$^{40}$Ca scattering cross-section, experimental data are available in the low-energy region between 15 and 100~MeV. As illustrated in Fig.~\ref{fig:2s1d}, all models except the Nakano $\textit{et. al.}$ model provides a reasonable description of the experimental data. Among these, the GEANT4 predictions show the smallest deviations from the measurements, especially for the high-energy range.

The next row of the nuclear periodic table corresponds to nuclei with valence nucleons occupying the $1f_{7/2}$ shell. Eight elements in this row occupy this state, with closing shells magic-number nuclei $^{59}$Ni. Due to the limited availability of experimental p-N inelastic cross-section data for this row, the present analysis is restricted to $^{47.9}$Ti, $^{51}$V, $^{56}$Fe, $^{59}$Co, and $^{59}$Ni, for which experimental proton--nucleus inelastic cross-section data are available.
For p-$^{47.9}$Ti and p-$^{51}$V scattering cross-section, experimental data are available in two distinct energy regions, namely below 100~MeV and between 4 and 10~GeV. 
The low-energy data for both have higher uncertainty than the high-energy measurements (Fig.~\ref{fig:1f7/2}-(a,b)). 
All the models show consistent results with the experimental measurement. Particularly, Letaw $\textit{et. al.}$ model and our model exhibit the smallest deviations from the experimental data among all the models considered. 
p-$^{56}$Fe and p-$^{59}$Co scattering cross-sections are measured for a small energy range up to energies below 1~GeV. All the models are showing reasonable agreement with the data for both cases. 
In the case of p-$^{59}$Ni scattering cross-section, only a few experimental data points are available. None of the models is able to reproduce the data satisfactorily; however, our model consistently shows smaller deviations from the experimental values compared to the other models, indicating a relatively improved performance. More experimental measurement is needed to shed light for Nickel target. 
The next row of the nuclear periodic table is with configuration $2p + 1f_{5/2}$ shells closed by the magic-number nuclei $^{91}$Zr. Out of twelve elements in this row, the present analysis is restricted to $^{63.5}$Cu, $^{65}$Zn, and $^{73}$Ge based on the availability of experimental measurements.
p-$^{63.5}$Cu scattering data are available over a wide energy range extending from 15~MeV up to 100~GeV, Fig.~\ref{fig:2p1f5/2}. The high-energy data \cite{Bobchenko:1979hp} are more precise than low energy measurement. 
As shown in Fig.~\ref{fig:2p1f5/2}-(a), because of the lack of data below approximately 70~MeV makes it difficult to draw firm conclusions regarding the precise position and magnitude of the low-energy peak.  
At higher energies, the saturation value predicted by our model deviates from the experimental data by approximately 4\%, while the GEANT4 model predicts more accurately.

For p-$^{65}$Zn scattering, These high-energy data, taken from Ref.~\cite{Grchurin1985}, are associated with relatively large experimental uncertainties, Fig.~\ref{fig:2p1f5/2}-(b).  
For the GeV range, most models fall within the experimental error bars. Our model shows relatively better prediction to the date at low-energy for both $^{65}$Zn and $^{73}$Ge,Fig.~\ref{fig:2p1f5/2}-(b,c).
The next row of the nuclear periodic table is the $1g_{9/2}$ shell with the magic-number nuclei $^{118.7}$Sn. Due to the limited availability of experimental the present analysis is restricted to $^{108}$Ag, $^{112.4}$Cd, and $^{118.7}$Sn, for which a small number of experimental p-N inelastic cross-section data are available.
Our model prediction shows a consistent result at high energy saturation; however, it fails to predict the low energy region satisfactorily for $^{118.7}$Sn. A quantitative comparison using the relative errors are listed in Table~\ref{tab:RE}, indicating that our model yields the smallest deviation among all the models considered. 


Similarly, Fig.\ref{fig:h11/2} shows cross-section for the measured elements of the $3s + 2d + 1g_{7/2} + 1h_{11/2}$ shells with the doubly magic nucleus $^{208}$Pb. Only the low-energy measurements are available for the targets $^{127}$I, $^{140}$Ce, $^{159}$Tb, and $^{197}$Au, while $^{207.2}$Pb has a high-energy measurement with a high error bar. Almost all the models and GEANT4 simulations show consistent results except Letaw $\textit{et. al.}$, which deviates by a large amount at the low-energy for all five targets. We notice that,  Letaw $\textit{et. al.}$ is not suitable at low-energy regime for the high A target.    

The last row of the nuclear periodic table is $3s + 2d + 1g_{7/2} + 1h_{11/2}$ shells with the doubly magic nucleus $^{289}$Fi. Out of thirty-two nuclei of this configuration, only  p-$^{238}$U cross-section is measured experimentally as shown in Fig.\ref{fig:i13/2}.
In the low-energy region, the Letaw $\textit{et. al.}$ model follows an almost constant trend for higher A nuclei, which is inconsistent with the experimental data. 

Out of the $33$ measured target nuclei, the $k$ shows a trend given by $k=\sum_{i=1} ^{3}\frac{\delta_i}{1 + exp(\theta_i (Z_{T_i} - Z))}$  as shown in Fig.\ref{fig:k_value}. Where  $\delta_i=\{-0.3, -0.6,-0.4\}$ and $\theta_i=\{3, 0.8,10 \}$, and the coulomb transition value  $Z_{T_i}=\{19, 28,70\}$ for $i=1,2,3$ respectively. 
\begin{table}[h!]
\centering
\caption{Values of $k$}
\label{tab:k_values}
\begin{tabular}{lc}
\hline\hline
Target &  $k$ \\
\hline
H, B & 0.9 \\
He & -2 \\
Li & 1.5\\
Be & 1.3 \\
N & 0.3 \\
F, Ne, Mg & 0.5 \\
Others & $\sum_{i=1} ^{3}\frac{\delta_i}{1 + exp(\theta_i (Z_{T_i} - Z))}$
\end{tabular}
\end{table}
The Coulomb term incorporates a strong repulsive interaction between the incident proton and the target nucleus at low incident energies.  The energy below $100~ MeV$, this effect dominates.
$k(Z)$ determines the location and height of the low-energy peak in the cross-section curve. The larger $k(Z)$ is, the greater the contribution of the Coulomb term, and hence the higher the peak. When $k(Z)$ is smaller or even negative, the suppression is greater, and the curve is moved accordingly. Thus, the parameter ensures the model mimics the target-dependent structure seen in experimental cross-sections.
The numerical value of $k(Z)$ is usually in the range from $-2$ to $1.5$. Such a comparatively small interval is enough to cover a broad range of nuclei, from light to heavy systems. A significant trend appears when the systematic behavior is analyzed for various atomic numbers: for lighter elements with $Z \leq 18$, the parameter $k(Z)$ is positive in most cases, i.e., the Coulomb term increases the peak in this region.
There are a few exceptional cases among the light nuclei, however. In the case of helium ($Z=2$), $k(Z)$ is negative. This is a peculiarity because helium, even though it is a light nucleus, has a cross-section behavior that is closer to that of heavy nuclei. Therefore, its Coulomb suppression could be described in a manner consistent with high-$Z$ behavior.
In addition to helium, certain other light nuclei like hydrogen, lithium, beryllium, nitrogen, fluorine, neon, and magnesium do not exhibit the same functional behavior. Their values are rather arbitrary and empirically fixed to reproduce the experimental data satisfactorily. 

For medium and heavy nuclei ($Z > 18$), the coefficient $k(Z)$ is always negative and shows a systematic step behavior reflecting a constant $k(Z)$ value for a group of elements, Fig.\ref{fig:k_value}. We can see that for light nuclei, the value of $k$ is arbitrary, but as we move to heavier nuclei, the $k$ values follow a step trend. These steps have dropping points at $ Z=18, 23$, and $70$.
The Coulomb term not only enhances the model's ability to account for low-energy scattering processes 
but also reflects that the cross-section is heavily convoluted by the interplay between the attractive nuclear forces and the repulsive Coulomb interaction, and the latter being particularly prominent below 100 MeV.
Such a Coulomb term could be considered as a generalization of the modification for beryllium in  Ref.\cite{letaw1983}, to accommodate a wider range of elements. 

Relative error, defined by Eq.\ref{eq:RE}, is used in this paper to compare how close the different model results are to the experimental data. A smaller value of relative error means the model is giving results that are closer to the measured data. In most cases, when the relative error is less than one, it can be taken as a sign that the model agrees well with the experimental results. The relative error is defined as
\begin{equation}\label{eq:RE}
    RE = \frac{1}{N} \sum \frac{(d_i - C_i)^2}{E_i^u + E_i^d},
\end{equation}
where,$N$ denotes the total number of data points,$d_i$ represents the value of the$i^{\text{th}}$ experimental data, and$C_i$ is the corresponding value obtained from the proposed formula. The terms$E_i^u$ and$E_i^d$ refer to the upper and lower bounds of the error bar for the$i^{\text{th}}$ experimental value, respectively \cite{nakano2021}.
 We compare the relative error with four popular models and GEANT4, the comparison, summarized in Table-\ref{tab:RE}, clearly demonstrates that the proposed model is more consistent than the existing models and GEANT4.

\subsection{Cosmic Rays Lifetime}
In the lifetime of a cosmic ray (CR) nucleus undergoing nuclear interactions with the interstellar medium (ISM), we first estimate the rate at which such interactions occur. The fundamental quantity of interest is the \textit{interaction rate}, which quantifies the number of collisions a cosmic ray nucleus experiences per unit time as it traverses a medium composed of target nuclei (e.g., hydrogen or helium atoms in the ISM).
Consider a CR nucleus of speed $v$, moving through a homogeneous medium of stationary target particles with number density $n$. In a small time interval $\Delta t$, the CR nucleus travels a distance $ v\Delta t$. The volume it sweeps through during this interval can be visualized as a cylinder with base area equal to the effective interaction cross-sectional area $\sigma(E)$, and height $v \Delta t$. The number of target particles within this volume is given by $n \cdot \sigma \cdot v \Delta t$. Each of these target particles represents a potential site for interaction. Therefore, the number of interactions occurring in time $\Delta t$ is $\Delta N = n \sigma(E) v \Delta t$, and the interaction rate $\Gamma$, defined as the number of interactions per unit time\cite{YAN2020121712}, is
\[
\Gamma = \frac{\Delta N}{\Delta t} = n \sigma(E) v.
\]
This equation is fundamental in kinetic theory and widely applicable in nuclear and particle transport physics. It implies that the interaction rate increases with target density, the effective cross-section, and the speed of the incident particle. If the medium contains multiple species of target particles (e.g., hydrogen and helium in the ISM), and each species $i$ has number density $n_i$ and interaction cross-section $\sigma_i(E)$, the total interaction rate becomes a sum over all species
\[
\Gamma = \sum_i n_i \sigma_i(E) v.
\]
For relativistic cosmic rays, the speed $v$ is conveniently expressed in terms of $\beta \equiv v/c$, where $c$ is the speed of light.
For a CR nucleus propagating through the ISM, the total interaction rate is the sum of contributions from hydrogen ($H$) and helium ($He$) nuclei
\[
\Gamma_{\text{total}} = \Gamma_H + \Gamma_{He} = \sigma_{\text{inel}}^{pZ}(E) \, n_H \, \beta c + \sigma_{\text{inel}}^{He}(E) \, n_{He} \, \beta c,
\]
where$n_H$ and$n_{He}$ are the number densities of ISM hydrogen and helium, respectively. The cross-sections  
$\sigma_{\text{inel}}^{pZ}(E)$ and$\sigma_{\text{inel}}^{He}(E)$ are take form Eq.(\ref{eq:loweq}). For interactions between CR nuclei and ISM protons ($pZ$), the improved cross-section formula is given by
\begin{align}
\sigma^{pZ}_{inel}((E))=&47.4A_{CR}^{0.676}[1-0.045cos(0.42 A_{CR} ^ {0.5})+0.000018A_{CR}^{1.63}] \\ \nonumber
&\times\left[(1 - 0.3 * exp(-E/3742) * \sin{(68*E ^ {-0.83})}) + \frac{1}{2} * exp(-E/134)\right]\\ \nonumber
&\times  \left[1 + k \cdot \exp{\left(\frac{-E}{24}\right)}\right].
\end{align}
Similarly, for interactions with ISM helium nuclei, the modified cross-section is expressed as
\begin{align}
\sigma^{\alpha Z}_{inel}(E)=&47.4[4^{0.388} +A^{0.388}_{CR}-1]^2][1-0.045cos(0.42 A_{CR} ^{0.5})+0.000018A_{CR}^{1.63}] \\ \nonumber 
&\times \left[(1 - 0.3 * exp(-E/3742) * \sin{(68*E ^ {-0.83})})
+ \frac{1}{2} * exp(-E/134)\right]\\ \nonumber
&\times  \left[1 + k \cdot \exp{\left(\frac{-E}{24}\right)}\right].
\end{align}
The total interaction rate$\Gamma_{\text{total}}$ is thus the sum of contributions from both hydrogen and helium, weighted by their respective cross-sections and densities. The lifetime$t_{\text{nuclear}}$ of the CR nucleus, defined as the mean time before a nuclear interaction occurs, is the inverse of the total interaction rate\cite{Lacki2012},
\begin{equation}
t_{\text{nuclear}} = \frac{1}{\Gamma_{\text{total}}} = \left[ \left( \sigma_{\text{inel}}^{pZ}(E) n_H + \sigma_{\text{inel}}^{\alpha Z}(E) n_{He} \right) \beta c \right]^{-1}.
\end{equation}
Equivalently, one may express the corresponding mean free path
D, i.e. the average distance travelled by a CR before undergoing an interaction, as
\begin{equation}
D = \left[ \left( \sigma_{\text{inel}}^{pZ}(E) n_H + \sigma_{\text{inel}}^{\alpha Z}(E) n_{He} \right)  \right]^{-1}.
\end{equation}
Where the values of $n_H$ and $n_{He}$ are taken from Ref.\cite{strong2007cosmic}. Our model result for the mean distance travelled by cosmic rays before interaction at 20 MeV is calculated to be $D=1.120 \times 10^7$ light years, which shows a deviation of around 20$\%$ from both the Shen's model prediction $D=1.374 \times 10^7 $ light years and the Letaw $\textit{et. al.}$ result $ D=1.311 \times 10^7 $ light years. The possible cause of the deviation is reflected in Fig.\ref{fig:1s}, where our model gives better predictions to the experimental data at low-energy peak for both $^2$H and $^4$He.  It indicates that our model produces values suggesting improved accuracy relative to the Letaw $\textit{et. al.}$ and Shen's model. 

\subsection{Antiproton Simulation}
In the Mu2e experiment at Fermilab\cite{bartoszek2015mu2etechnicaldesignreport}, the primary goal is to search for the neutrinoless muon-to-electron conversion process ($\mu^{-} \rightarrow e^{-}$)in the field of an aluminium nucleus. The experimental signature of this rare process is a monoenergetic electron with an energy of 104.97 MeV for the aluminium stopping target. However, a critical challenge in the detection of this signal arises from several potential background sources that can mimic or obscure the conversion electron. One such notable background arises from antiprotons produced when the high-intensity 8.9 GeV proton beam strikes the production target.
These antiprotons, if not sufficiently absorbed or diverted, may reach the stopping target region, annihilate, and produce high-energy electrons, some of which lie in the vicinity of the signal region. Therefore, an accurate estimation of the antiproton yield per proton on target (POT) is essential for predicting and mitigating this background contribution.
The antiproton production rate per POT is commonly expressed as
\begin{equation}
\frac{N_{\bar{p}}^{PT}}{\text{POT}} = \frac{\sigma_{\bar{p}}}{\sigma_{\text{inelastic}}} \cdot \frac{N_{\text{inelastic}}}{N_{\text{POT}}}.
\end{equation}
Where $\sigma_{\bar{p}}$ is the total antiproton production cross-section, $\sigma_{\text{inelastic}}$ is the total inelastic cross-section for protons on the target material, $N_{\text{inelastic}} / N_{\text{POT}}$ $\approx{0.792}$ is the probability obtained from Monte Carlo simulation \cite{mu2ecollaboration2022mu2erunisensitivity} and ${N_{\bar{p}}^{PT}}$ is antiprotons produced in the production target.

The empirical model by Letaw $\textit{et. al.}$ provides a widely cited value for the inelastic cross-section which is $\sigma_{\text{inelastic}} = 1707$ mb, yielding an antiproton production rate $N_{\bar{p}}^{PT}= 6.547 \times 10^{-5}$. While our empirical model result is  $\sigma_{\text{inelastic}} = 1689$ mb with 
$N_{\bar{p}}^{PT}=6.620 \times 10^{-5}$. This model employs a more refined parametrization of inelastic cross-sections from updated experimental data and improved theoretical treatments.
Furthermore, the Monte Carlo N-particle transport code(MCNP), a simulation toolkit benchmark in high-precision nuclear interaction modeling, provides $\sigma_\text{inelastic} = 1517$ mb and the corresponding antiproton production is  $N_{\bar{p}}^{PT}= 7.371 \times 10^{-5}$ \cite{MCNP_2012}. MCNP provides a detailed transport simulation with high accuracy and incorporates realistic geometric and material configurations, making its results a widely accepted reference.
Our model’s prediction shows a deviation of around 11$\%$ (larger) than MCNP and consistent with Letaw $\textit{et. al.}$’s value ( with around 1$\%$ deviation). 

\section{Conclusion}\label{conclusion}
In this paper, we propose an empirical formula to estimate the inelastic cross-section for proton–nucleus scattering of various target nuclei over a wide range of interaction energy. Compared to the existing models, the proposed model is relatively simple with less numbers of parameters and provides consistent results for both light and heavy nuclei in the energy range of 15 MeV to 1 TeV. The model is based on the assumption that the proton inelastic cross-section depends on the energy of the incident proton ($E$), atomic mass ($A$), and atomic number ($Z$) of the target.
However, at higher energies, the cross-section tends to saturate at a constant value, which differs for different target nuclei based on the mass number $A$. The parameters of the model are determined by the fitting of the scattering cross-section saturation value data measured experimentally.  
While at the lower energy regime, the experimental measurement for the cross-section shows irregular behavior with around 60$\%$ higher peak at $E<200$ MeV and a drop of around 15$\%$ around 1 GeV. The cross-section is factorized into three terms. One is the high-energy empirical formula of mass number dependency, which ensures the saturation value. Others are energy dependency term and a Coulomb term. The low-energy variation of this proposed empirical formula is modeled, incorporating the dominant Coulomb effect that captures the behavior and peak intensity. The parameters of these two terms are fixed by the fitting of the p-$^{12}$C and p-$^{27}$Al inelastic cross-section data measured in several experiments. The model prediction for the proton scattering cross-section data of various target nuclei and a detailed comparison with the existing models are presented. This proposed empirical formula shows relatively better precision as well as qualitative agreement with the experimental inelastic cross-section data for most of the 33-target nuclei measured. For completeness, a GEANT4 simulation for all these targets are also included in the numerical prediction and comparison.
For heavy elements, Letaw $\textit{et. al.}$ model shows a huge overestimated peak in the low energy region, whereas our model shows consistency with the data. Our model result for cosmic ray lifetime is estimated by calculating the average distance travelled by a cosmic ray, which shows around 20$\%$ improvement at low energy than Shen's and Letaw $\textit{et. al.}$ models prediction.   This model results for the antiproton production rate per POT shows an agreement with the $\textit{et. al.}$ prediction but deviates by around 10.8$\%$ than the MCNP simulation result.  

\begin{acknowledgments}
Work of TM is supported by the Anusandhan National Research Foundation (ANRF), Department of Science and Technology, Government of India and Science and Engineering Research Board (SERB) through the SRG (Start-up Research Grant) of File No.
SRG/2023/001093.

\end{acknowledgments}

\appendix

\section{}\label{AppA}
Our high-energy formula Eq.(\ref{eq:highour}) fitted with the experimental data corresponding to the proton incident energy ranging from 5 to 9 GeV, is presented in Table \ref{table: High energy Exp. data}, which is taken from Ref.~\cite{Bobchenko:1979hp}.
\begin{table}[h]
\begin{center}
\begin{tabular}{ |c|c|c|c|c|c|c|c|c|c|c|c|c|c|c|c|c|c|c| } 
 \hline
Element&  $^9$Be& $^{10.8}$B& $^{12}$C&  $^{19}$F&  $^{24.3}$Mg& $^{27}$Al& $^{32.1}$S&  $^{40.1}$Ca& $^{48.9}$Ti \\
\hline
Cross-section(mb)& $209\pm3$& $235\pm4$& $251\pm2$& $351\pm5$& $422\pm5$& $456\pm7$& $514\pm6$& $603\pm6$& $683\pm8$\\
 \hline
\end{tabular}
\end{center}
\end{table}
\vspace{-.8cm}
\begin{table}[h]
\begin{center}
\begin{tabular}{ |c|c|c|c|c|c|c|c|c|c|c|c|c|c|c|c|c|c|c| } 
 \hline
Element&$^{50.9}$V&  $^{55.8}$Fe& $^{63.5}$Cu& $^{92.9}$Nb& $^{112.4}$Cd& $^{118.7}$Sn& $^{180.9}$Ta& $^{207.2}$Pb& $^{238}$U\\
\hline
Cross-section(mb)& $714\pm8$& $760\pm8$& $831\pm8$& $1070\pm11$& $1215\pm12$ & $1255\pm13$& $1666\pm19$& $1859\pm16$& $2090\pm45$\\

 \hline
\end{tabular}
\end{center}
\caption{Condensation of the Total Inelastic Cross Section Data. \cite{Bobchenko:1979hp}}
\label{table: High energy Exp. data}
\end{table}

\begin{table}[h]
\begin{center}
\begin{tabular}{|c|c|c|c|c|c|c|c|}
\hline
 Z & Elements  &  Letaw   &  Shen  &  Tripathi  &  Nakano \footnote{For Nakano $\textit{et. al.}$'s model, we have taken the energy range of 15 MeV to 1 GeV. For others, it is 15 MeV to 1 TeV.} & GEANT4 & Our Model \\
 \hline 
1 & $^2$H           &1.75&1.64&3.57 & -&8.00&0.59 \\
\hline 
2&$^4$He            &2.03&3.38&1.98&4.73&0.50&1.45\\
\hline 
3&$^6$Li            &1.91&2.36&0.20&4.87&1.53&0.83\\
\hline 
4&$^9Be$            &0.50&3.32&1.35&4.43&0.42&0.81\\
\hline
5&$^{10.8}$B             &1.69&1.78&1.57&2.17&0.62&0.53\\
\hline 
6&$^{12}$C             &1.12&1.15&0.78&0.86&0.66&0.72\\
\hline 
7&$^{14}$N             &0.98&0.69&0.43&0.67&0.29&0.27\\
\hline 
8&$^{16}$O             &1.32&0.80&0.39&0.66&0.31&0.41\\
\hline 
9&$^{19}$F             &0.79&1.17&0.62&1.72&1.22&0.43\\
\hline 
10&$^{20}$Ne           &2.65&2.19&1.06&2.20&1.90&0.71\\
\hline 
11&$^{23}$Na           &0.39&0.27&0.19&0.22&0.28&0.27\\
\hline 
12&$^{24}$Mg           &4.64&3.96&0.93&3.49&3.31&2.10\\
\hline 
13&$^{27}$Al           &0.73&1.08&0.80&0.62&0.75&0.66\\
\hline 
14&$^{28}$Si           &2.05&2.38&2.28&0.37&1.27&1.70\\
\hline 
18&$^{40}$Ar           &1.74&1.64&1.09&0.77&1.85&1.26\\
\hline 
20&$^{40}$Ca           &0.93&0.69&1.06&2.16&0.54&0.69\\
\hline 
22&$^{47.9}$Ti           &0.41&2.84&1.57&0.63&0.89&0.48\\
\hline 
23&$^{51}$V           &0.49&2.94&1.58&0.83&0.94&0.59\\
\hline 
26&$^{56}$Fe           &0.84&0.61&0.70&0.44&0.69&0.50\\
\hline 
27&$^{59}$Co          &1.15&0.85&0.89&0.78&0.69&0.61\\
\hline 
28&$^{59}$Ni          &2.23&1.76&1.41&1.43&1.69&1.16\\
\hline 
29&$^{63.5}$Cu           &0.91&2.43&1.97&0.80&0.82&0.81\\
\hline 
30&$^{65}$Zn           &0.57&0.40&0.19&1.58&0.92&0.27\\
\hline 
32&$^{73}$Ge           &0.84&1.45&1.52&1.12&0.63&0.99\\
\hline 
47&$^{108}$Ag           &1.36&1.23&1.72&1.00&0.92&0.81\\
\hline 
48&$^{112.4}$Cd           &0.55&1.90&2.79&0.86&0.67&0.57\\
\hline 
50&$^{118.71}$Sn           &2.49&2.34&2.85&2.72&1.47&1.20\\
\hline 
53&$^{127}$I          &1.00&0.88&1.59&0.90&0.42&1.33\\
\hline 
58&$^{140}$Ce           &5.26&2.03&0.48&0.87&1.05&1.17\\
\hline 
65&$^{159}$Tb           &2.59&0.62&0.18&0.37&0.55&0.73\\
\hline 
79&$^{197}$Au           &3.52&0.71&0.69&0.56&0.74&0.65\\
\hline 
82&$^{207.2}$Pb           &1.95&1.01&2.08&0.90&0.75&0.72\\
\hline 
92&$^{238}$U           &0.90&1.09&2.54&0.44&0.73&0.71\\
\hline 
\end{tabular}
\end{center}
\caption{Relative error (RE) Comparison for all 33 elements}
\label{tab:RE}
\end{table}

Table~\ref{table: experimental data citations} lists the references of the experimental data shown in Fig.\ref{fig:1s}-\ref{fig:i13/2}, and compared with the model predictions.
The various selected datasets represent a large variety of nuclei and energies; therefore, they provide a consistent and reliable basis for the benchmark of the present calculations.
\begin{table}[h]
\begin{center}
\begin{tabular}{|c|c|c|c|c|c|c|c|c|}
\hline
Z & Element & Data Source (Reference) & Z & Element & Data Source (Reference)& Z & Element & Data Source (Reference) \\
\hline
1  & H   & \cite{carlson1996} &12 & Mg  & \cite{carlson1996}&30 & Zn  & \cite{carlson1996,Grchurin1985}\\
2  & He  & \cite{carlson1996, PhysRevD.23.1895, Velichko1984}&13 & Al  & \cite{carlson1996,Bobchenko:1979hp,denisov1973absorption,fumuro1979dependence, Grchurin1985}& 32 & Ge  & \cite{carlson1996}\\
3  & Li  & \cite{carlson1996, Grchurin1985}&14 & Si  & \cite{carlson1996,Bobchenko:1979hp}& 47 & Ag  & \cite{carlson1996,goloskie1962measurement} \\
4  & Be  & \cite{carlson1996,Bobchenko:1979hp,denisov1973absorption,Grchurin1985,Slaus:1975zz}&18 & Ar  & \cite{carlson1996,de1982multiparticle,L}& 48 & Cd  & \cite{carlson1996,Bobchenko:1979hp}\\
5  & B   & \cite{carlson1996,Bobchenko:1979hp}&20 & Ca  & \cite{carlson1996}& 50 & Sn  & \cite{carlson1996,Bobchenko:1979hp,denisov1973absorption}\\
6  & C   & \cite{carlson1996,Bobchenko:1979hp,Grigorov:1970xu,Grchurin1985}&22 & Ti  & \cite{,carlson1996,Bobchenko:1979hp}&  53 & I   & \cite{carlson1996}\\
7  & N   & \cite{carlson1996}&23 & V   & \cite{carlson1996,Bobchenko:1979hp}& 58 & Ce  & \cite{carlson1996} \\
8  & O   & \cite{carlson1996,Barshay:1974sc} &26 & Fe  & \cite{carlson1996}&65 & Tb  & \cite{carlson1996}\\
9  & F   & \cite{carlson1996,Bobchenko:1979hp}&27 & Co  & \cite{carlson1996}& 79 & Au  & \cite{carlson1996,abegg1979}\\
10 & Ne  & \cite{carlson1996}&28 & Ni  & \cite{carlson1996}& 82 & Pb  & \cite{carlson1996,Bobchenko:1979hp,denisov1973absorption,institute1966proceedings,Grchurin1985,Lattes1965}\\
11 & Na  & \cite{carlson1996}&29 & Cu  & \cite{carlson1996,Bobchenko:1979hp,denisov1973absorption}& 92 & U   & \cite{carlson1996,Bobchenko:1979hp,denisov1973absorption,Grchurin1985}\\
\hline
\end{tabular}
\end{center}
\caption{Elements and corresponding references for experimental data sources used }
\label{table: experimental data citations}
\end{table}

\begin{figure}[h]
    \centering
    \includegraphics[width=0.48\linewidth, trim=100 240 100 240, clip]{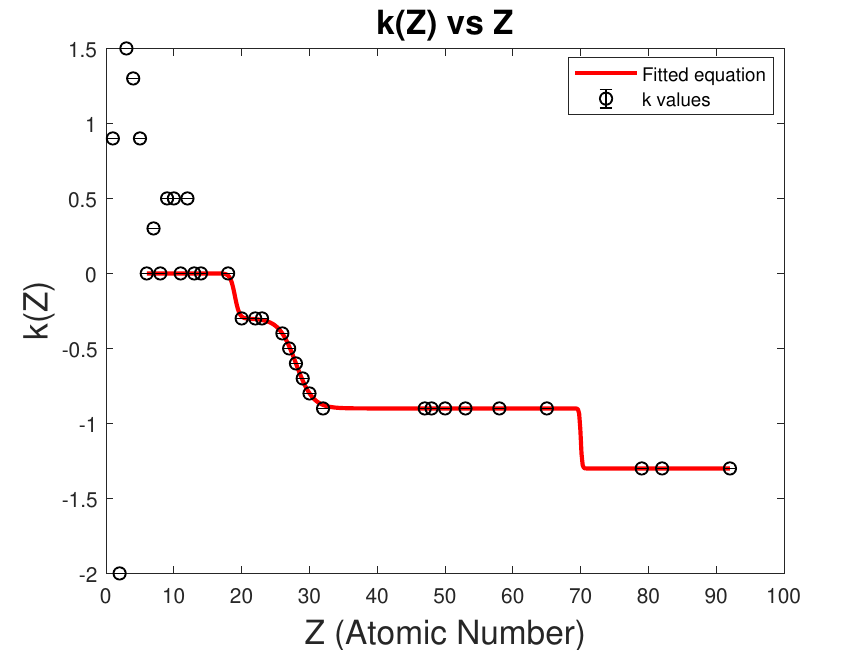}
    \caption{Coulomb parameter behaviour }
    \label{fig:k_value}
\end{figure}

\begin{table}[h]
\centering
\renewcommand{\arraystretch}{1.2}
\setlength{\tabcolsep}{6pt}

\begin{tabular}{|c|c|c|c|c|c|c|c|c|c|}
\hline
H(1) & He(2) & Li(3) & Be(4) & B(5) & C(6) & N(7) & O(8) & F(9) & Ne(10) \\
\hline
72.9 & 117.4 & 155 & 206.6 & 235 & 253.2 & 282.6 & 310.9 & 351.7 & 364.9 \\
\hline
Na(11) & Mg(12) & Al(13) & Si(14) & Ar(18) & Ca(20) & Ti(22) & V(23) & Fe(26) & Co(27) \\
\hline
403.5 & 416.1 & 453 & 465 & 600.9 & 600.9 & 682.9 & 713.8 & 762 & 790 \\
\hline
\end{tabular}

\vspace{0.4cm}

\begin{tabular}{|c|c|c|c|c|c|c|c|c|}
\hline
Ni(28) & Cu(29) & Zn(30) & Ge(32) & Ag(47) & Cd(48) & Sn(50) \\
\hline
790 & 831 & 844.4 & 913.6 & 1181.9 & 1212.8 & 1256.3 \\
\hline
I(53) & Ce(58)&Tb(65) & Au(79) & Pb(82) & U(92) &  \\
\hline
1312.4 & 1398.9&1524.1 & 1782.5 & 1861.3 & 2089.1 & \\
\hline
\end{tabular}
\caption{Asymptotic inelastic cross-section values (in mb) of our model at high energies for different elements.}
\label{table:asymptotic}
\end{table}

\bibliographystyle{unsrt}   
\bibliography{WD_Zeta_v3} 

\end{document}